\documentclass[prx,aps,amsmath,amssymb,showpacs,twocolumn]{revtex4-2}
\usepackage{graphicx,amsmath,amssymb,color}
\usepackage{nicefrac}
\usepackage{physics}
\usepackage{multirow, makecell, ragged2e}
\usepackage[titletoc,title]{appendix}

\newcommand{\be}{\begin{equation}}
\newcommand{\ee}{\end{equation}}

\newcommand{\ba}{\begin{eqnarray}}
\newcommand{\ea}{\end{eqnarray}}

\renewcommand{\vec}[1]{\mbox{\boldmath$#1$}}

\def\beq{\begin{eqnarray}}
\def\eeq{\end{eqnarray}}
\newcommand{\mi}{{i}}

\newcommand{\elliptic}[5][\scriptstyle]{\vartheta\left[\begin{array}{c}{{#1 #2}}\\{#1 #3}\end{array}\right]\left(#4\middle|#5\right)}

\newcommand{\sh}{\mathcal{S}}

\makeatletter
\newcommand*{\rom}[1]{\expandafter\@slowromancap\romannumeral #1@}
\makeatother

\begin{document}

\title{Bardeen-Cooper-Schrieffer pairing of composite fermions}

\author{Anirban Sharma, Songyang Pu and J. K. Jain}
\affiliation{Department of Physics, 104 Davey Lab, Pennsylvania State University, University Park, Pennsylvania 16802,USA}

\date{\today}

\begin{abstract} 
Topological pairing of composite fermions has led to remarkable ideas, such as excitations obeying non-Abelian braid statistics and topological quantum computation. We construct a $p$-wave paired Bardeen-Cooper-Schrieffer (BCS) wave function for composite fermions in the torus geometry, which is a convenient geometry for formulating momentum space pairing as well as for revealing the underlying composite-fermion Fermi sea.  Following the standard BCS approach, we minimize the Coulomb interaction energy at half filling in the lowest and the second Landau levels, which correspond to filling factors $\nu=1/2$ and $\nu=5/2$ in GaAs quantum wells, by optimizing two variational parameters that are analogous to the gap and the Debye cut-off energy of the BCS theory. Our results show no evidence for pairing at $\nu=1/2$ but a clear evidence for pairing at $\nu=5/2$. To a good approximation, the highest overlap between the exact Coulomb ground state at $\nu=5/2$ and the BCS state is obtained for parameters that minimize the energy of the latter, thereby providing support for the physics of composite-fermion pairing as the mechanism for the $5/2$ fractional quantum Hall effect. We discuss the issue of modular covariance of the composite-fermion BCS wave function, and calculate its Hall viscosity and pair correlation function. By similar methods, we look for but do not find an instability to $s$-wave pairing for a spin-singlet composite-fermion Fermi sea at half-filled lowest Landau level in a system where the Zeeman splitting has been set to zero.
\end{abstract}

\maketitle

\section{Introduction}

The fractional quantum Hall effect (FQHE)~\cite{Tsui82} has proved a treasure trove of exotic emergent phenomena. A striking example is the FQHE at filling factor $\nu=5/2$~\cite{Willett87,Pan99}, which corresponds to half filled second Landau level (LL) in GaAs quantum well systems. The most promising theoretical explanation of this state~\cite{Moore91,Read00} passes through a succession of remarkable emergences: First is the emergence of composite fermions (CFs), namely electrons carrying two quantized vortices, which arise as a result of the repulsive interaction between electrons~\cite{Jain89,Jain07,Halperin20}. Composite fermions experience no effective magnetic field at half filling, and attempt to form a CF Fermi sea (CFFS), in analogy to the CFFS at $\nu=1/2$ in the lowest LL (LLL)~\cite{Halperin93,Willett93,Kang93,Goldman94,Smet96,Halperin20b,Shayegan20}. The CFFS in the second LL (SLL), however, is unstable to a topological $p$-wave pairing of fully spin polarized composite fermions, which opens a gap and thus produces a FQHE. Furthermore, this paired state is predicted to give birth to its own new emergent particles, namely Majorana particles obeying non-Abelian braiding statistics~\cite{Moore91,Read00}. These are interesting in their own right and have also generated exciting proposals for topological quantum computation~\cite{Nayak08}. The past three decades have seen an intense theoretical and experimental investigation of the ``$5/2$ state," which has lent nontrivial support to certain aspects of the above-outlined physical mechanism for the 5/2 FQHE. Moore and Read (MR) proposed an ansatz wave function for the paired CF state~\cite{Moore91}, which has a lower energy than the CFFS~\cite{Park96} and a significant overlap with the exact Coulomb ground state for small systems~\cite{Morf98}. Furthermore, numerical calculations indicate that the CFFS in the second Landau level (LL) is unstable to Cooper pairing~\cite{Scarola00b}. More recently, it has been shown~\cite{Balram18} that a wave function belonging to the parton class~\cite{Jain89b} also describes topological superconductivity of composite fermions and  provides a comparably decent quantitative account of the exact Coulomb state.  Experimentally, convincing evidence exists that the 5/2 state in the SLL is fully spin polarized~\cite{Tiemann12,Stern12,Eisenstein17,Hossain18b}, which is a necessary condition for topological $p$-wave superconductivity. The appearance of a CFFS at $\nu=5/2$ at either elevated temperatures~\cite{Willett02} or at nearby filling factors~\cite{Hossain18b} supports the notion that the 5/2 state arises from an instability of the CFFS. Furthermore, the thermal Hall conductance of the 5/2 state has been found to be half quantized~\cite{Banerjee17b}, as expected from topological superconductivity, although its value is inconsistent with the expectation from the MR state or its hole conjugate.

Even though the MR wave function can be readily seen to describe pairing of composite fermions, it is not expressed in the standard Bardeen-Cooper-Schrieffer (BCS) form.  There are several motivations to construct a CF-BCS wave function. For one thing, the MR (or the parton) wave function does not contain any variational parameters that would allow one to optimize the pair wave function. (The absence of variational parameters is a rather ubiquitous feature of the CF theory, but often, especially in the LLL, the parameter-free wave functions turn out to be such accurate representations of the Coulomb ground states that the lack of variational parameters may be seen as a virtue rather than a shortcoming.) Second, a BCS wave function should clarify how the paired state evolves out of the CFFS. Finally, the BCS framework can in principle be used to study pairing of other symmetry types that might be relevant at other filling factors.

We construct in this article a BCS wave function for composite fermions. A significant aspect of our work is that we employ the torus geometry, which is the most natural geometry for formulating pairing in the momentum space as well as for revealing the underlying CFFS. We follow the standard procedure of the composite-fermion theory~\cite{Jain89,Jain07,Halperin20} by first constructing the BCS pair wave function for electrons and then composite-fermionizing it by vortex attachment. There are several technical obstacles that must be overcome, however. In particular, the standard Jain-Kamilla (JK) projection method \cite{Jain97,Jain97b} does not preserve the quasi-periodic boundary conditions. We show how a suitable modification of this method accomplishes the goal and produces legitimate LLL wave functions. (We note that it is convenient to work within the LLL subspace; the SLL physics is simulated in the LLL through an effective interaction that has the same Haldane pseudopotentials~\cite{Haldane83} in the LLL as the Coulomb interaction does in the SLL.) This wave function has two variational parameters, analogous to the gap parameter and the Debye cut-off of the BCS theory, and reduces to the CFFS in one limit.  We calculate the Coulomb energy of this wave function as a function of these parameters. We find that at $\nu=5/2$, the lowest energy is obtained for a non-zero gap, indicating a pairing instability of the CFFS. No pairing instability is found at $\nu=1/2$; here the minimum energy is produced by the CFFS state.

There has been important previous work along this direction. M\"oller and Simon~\cite{Moller08}  implemented a CF-BCS wave function in the spherical geometry. They showed that a good approximation could be found for the exact Coulomb ground state as well as the MR wave function for appropriate choices of parameters. More recently, Wagner {\em et al.}\cite{Wagner21} have considered $s$-wave pairing in quantum Hall bilayers, wherein an electron-flux composite fermion in one layer and a hole-flux composite fermion in the other layer form pairs and condense. Yutushui and Mross~\cite{Yutushui20} have shown how to study CF pairing for large scale systems. All of these studies have employed the spherical geometry. 

We note that even though we follow the convention of using the terminology ``$5/2$ state," which refers to the FQHE state at half-filled SLL in GaAs based quantum wells, completely analogous states occur at other half fillings in GaAs quantum wells (e.g. $7/2$), AlAs quantum wells, and bilayer graphene~\cite{Zibrov16,Li17}. 
Our work also applies to all of these states to the extent that corrections due to LL mixing and finite width may be neglected. 

While the actual state at $\nu=5/2$ is believed to be fully spin-polarized even in the absence of Zeeman energy, the state at $\nu=1/2$ is believed to be a spin-singlet CFFS in the limit of vanishing Zeeman energy. We investigate whether the $\nu=1/2$ CFFS is unstable to spin-singlet pairing of composite fermions when the Zeeman coupling is switched off. We find no signature of s-wave pairing for Coulomb interaction.

The paper is organized as follows. In Sec. II we review various previously known wave functions of composite fermions on a torus. In Sec. III we construct a BCS wave function for fully spin-polarized composite fermions with two variational parameters: a gap parameter and a momentum cutoff. It is shown that a modified JK projection preserves the quasiperiodic boundary conditions to produce legitimate LLL wave functions. The variational parameters are determined by energy minimization in Sec. IV. The CFFS is found to be the lowest energy state at $\nu=1/2$, indicating an absence of pairing instability. In contrast, at $\nu=5/2$, pairing of composite fermions leads to a lower energy. We compare the CF-BCS state with the exact Coulomb ground state as well as the MR wave function.
In Sec. V we use this method to look for a spin singlet pairing instability at $\nu=1/2$ in a system where the Zeeman energy is switched off. We calculate the Hall viscosity of the CF-BCS state in Sec. VI. the appendices provide various details omitted from the main text as well as a brief review of the numerical methods.

\section{Composite fermions on a torus}

The torus geometry, with a magnetic field perpendicular to its surface, was introduced for FQHE in the 1980s~\cite{Yoshioka83,Haldane85,Haldane85b}. We begin with a brief review of the various CF states on the torus, namely the Laughlin, Jain, MR and CFFS states, which introduces concepts and technical details that are necessary when we construct the CF-BCS wave function. 

The torus is represented by a parallelogram in the complex plane with periodic boundary conditions. The two sides of the parallelogram are given by $L$ and $L\tau$, where $L$ is taken to be along the real axis, and $\tau=\tau_1+i\tau_2$ is a complex number representing the modular parameter of a torus~\cite{Gunning62}.  The magnetic field is perpendicular to the parallelogram with $\vec{B}=-B\hat{z}$. We use the symmetric gauge with $\vec{A}=\frac{B}{2}(y,-x,0)$, and define the complex coordinates as $z=x+iy$. The single-particle wave functions are taken to satisfy the quasi-periodic boundary conditions along the two directions
\ba
t(L)\psi(z,\bar{z}) = e^{i\phi_1}\psi(z,\bar{z})  \\ \nonumber
t(L\tau)\psi(z,\bar{z}) = e^{i\phi_{\tau}}\psi(z,\bar{z})
\ea
where $t(L)$ and $t(L\tau)$ are the magnetic translation operators along the two edges of the parallelogram, whose general definition is 
\beq
t(\xi)= e^{-\frac{i}{2\ell^2}\hat{\vec{z}}\cdot (\vec{\xi} \times \vec{r})}T(\xi)
\eeq
 where $\ell=\sqrt{\hbar c/eB}$ is the magnetic length and $T(\xi)$ is the translation operator for a vector $\vec{\xi}=(\xi_x, \xi_y)$ defined as 
\beq
T(\xi) = e^{\xi\partial_z+ \bar{\xi}\partial_{\bar{z}}}
\eeq
with 
$\xi=\xi_x+i\xi_y$. The phases $\phi_{1}$ and $\phi_{\tau}$ specify the Hilbert space~\cite{Haldane85b}. The commutation relation $[t(L),t(L\tau)]=0$ requires the number of flux quanta through the torus, $N_{\phi}=L^2\rm{Im(\tau)}B/\phi_0$, to be an integer, where $\phi_0=hc/e$ is a flux quantum. 
The relations
\beq
t(L)e^{z^2-|z|^2 \over 4 \ell ^2} = e^{z^2-|z|^2 \over 4 \ell ^2} T(L)
\eeq
\beq
t(L\tau)e^{z^2-|z|^2 \over 4 \ell ^2} = e^{z^2-|z|^2 \over 4 \ell ^2}e^{-i\pi N_{\phi}(2z/L+\tau)} T(L\tau)
\eeq
will be useful below.
The many-particle wave function $\Psi$ with $N$ particles should satisfy the properties
\ba
t_j(L)\Psi(\{z_i\},\{\bar{z}_i\})= e^{i\phi_1}\Psi(\{z_i\},\{\bar{z}_i\})  \\ \nonumber
t_j(L\tau)\Psi(\{z_i\},\{\bar{z}_i\})= e^{i\phi_{\tau}}\Psi(\{z_i\},\{\bar{z}_i\})
\ea
where $t_j$ is the magnetic translation operator for the $j^{th}$ particle with $j=1,2,...,N$. 

We now briefly review some model wave functions that we use in our calculations. We assume absence of LL mixing and thus project the wave functions into the LLL. We simulate the physics of the second LL by mapping the Coulomb interaction to an effective interaction in the LLL \cite{Park98b, Scarola00}. 

{\it The Laughlin wave function}: The Laughlin wave function at filling factor $\nu=1/m$, which we use later, is written in the disk geometry as~\cite{Laughlin83} 
\be
\Psi^{\rm L}_{1/m}=\exp[-\sum_j|z_j|^2/4\ell^2]\prod_{i<j}(z_i-z_j)^m.
\ee
In torus geometry, the analogous wave function satisfying the periodic boundary conditions is given by \cite{Haldane85,Haldane85b,Pu20b}:
\begin{widetext}
\beq
\label{Laughlin}
{\Psi}^{\rm L}_{1/m,k_{\rm CM}}  = e^{\sum_i {z_i^2-\abs{z_i}^2 \over 4 \ell^2}}\Bigg \{ \vartheta
\begin{bmatrix}
{\phi_1\over 2\pi m}
 +{ k_{\rm CM} \over m} + {N-1 \over 2}\\
-{\phi_{\tau}\over 2\pi } + {m(N-1)\over 2}
\end{bmatrix}
\Bigg({mZ \over L} \Bigg |m \tau \Bigg) \Bigg \}  \prod_{i<j} \Bigg [ \vartheta 
\begin{bmatrix}
{1\over 2} \\ {1\over 2}
\end{bmatrix}
\Bigg( {z_i-z_j\over L}\Bigg | \tau \Bigg ) \Bigg ]^m ,
\eeq
\end{widetext}
where $Z=\sum_{i=1}^Nz_i$ is the center-of-mass (COM) coordinate, and $k_{\rm CM}=0,..,m-1$ is related to the COM momentum, defined through:
\beq
t_{\rm CM}\left(L/N_\phi\right) {\Psi}^{\rm L}_{1/m,k_{\rm CM}} =e^{i2\pi({\phi_1\over 2\pi m}
 +{ k_{\rm CM} \over m} + {N-1 \over 2})} {\Psi}^{\rm L}_{1/m,k_{\rm CM}}
\eeq
where 
\beq
t_{\rm CM}\left(L/N_\phi\right)\equiv \prod_{i=1}^Nt_i\left(L/N_\phi\right).
\eeq
 The $m$ values of $k_{\rm CM}$ refer to $m$ degenerate ground state wave functions. Here we use the Jacobi theta function with rational characteristics, defined as \cite{Mumford07}:
\beq
\vartheta  \begin{bmatrix}
a \\ b
\end{bmatrix}(z|\tau) =\sum_{n = -\infty}^{\infty} e^{i\pi(n+a)^2 \tau} e^{i2 \pi(n+a)(z+b)}.
\eeq
The factor $\prod_{i<j} \vartheta 
\begin{bmatrix}
{1\over 2} \\ {1\over 2}
\end{bmatrix}
\Bigg( {z_i-z_j\over L}\Bigg | \tau \Bigg )$ is analogous to $\prod_{i<j}(z_i-z_j)$ of the disk geometry \cite{AlvarezGaume86}. For the special case of $m=1$, Eq.~\ref{Laughlin} gives the wave function $\Psi_1$ for the filled LLL.

{\it The Jain wave functions}: For the ground and excited states at arbitrary filling factors $\nu=N/N_\phi$, where $N$ particles are exposed to $N_\phi$ flux quanta, the Jain wave functions are constructed as:
\beq
\label{Jain}
\Psi_{\nu^*/(2p\nu^*+1)} ={ P_{\rm LLL}} \Psi_{\nu^*} \Psi_1^{2p}
\eeq
where $P_{\rm LLL}$ is the LLL projection operator, and $\Psi_{\nu^*}$ is the many-particle wave function at filling factor $\nu^*=N/N_\phi^*$, with $N_\phi^*=N_\phi-2pN$. For integer values of $\nu^*=N/N_\phi^*=n$, $\Psi_{n}$ is a Slater determinant representing $n$ filled LLs, and the wave functions $\Psi_{n/(2pn+1)}$ correspond to the incompressible Jain states \cite{Jain89,Pu17}. 

The LLL projection can be accomplished in more than one way. A ``direct" projection~\cite{Dev92} is equivalent to expanding the wave function in the Slater determinant basis and retaining only the LLL part. Because the dimension of the Hilbert space grows exponentially with the system size, this projection can be accomplished only for relatively small systems. An alternative method is the JK projection~\cite{Jain97,Jain97b}, which allows treatment of much larger systems. The JK projection must be modified in the torus geometry, as shown in Ref.~\cite{Pu17}, which gives the explicit LLL projected form of the wave functions in the torus geometry. We note that the wave function in Eq.~\ref{Jain} is, in general, not an eigenstate of the COM momentum; Ref.~\cite{Pu17} has shown how all degenerate ground states with definite COM momenta can be obtained from it. 

{\it The CFFS wave function}: 
For the special case of $N_\phi^*=0$, $\Psi_{\nu^*}$ is a slater determinant of plane waves, ${\rm Det}\left[e^{i\vec{k}_n\cdot\vec{r}_m}\right]$, producing the CFFS wave function at filling $\nu=1/2$:
\beq
\Psi^{\rm CFFS}_{1/ 2,k_{\rm CM}} =  P_{\rm LLL} {\rm Det}[e^{i\vec{k}_n\cdot\vec{r}_m}]\Psi_{1/2,k_{\rm CM}}^{\rm L} 
\label{CFFS1}
\eeq
where we have used $\Psi_{1/2,k_{\rm CM}}^{\rm L}$ rather than $\Psi_1^2$ to ensure that the wave function has a well-defined COM momentum by choosing $k_{\rm CM}=0,1$. In the rest of the paper, we take $k_{\rm CM}=0$ and omit this subscript. 
The wave vectors $\vec{k}$'s that are allowed by the quasi-periodic boundary conditions for a torus are: 
\beq
\vec{k}_n = \left[n_1+{\phi_1\over 2\pi} \right]\vec{b_1} + \left[n_2+{\phi_\tau\over 2\pi} \right] \vec{b_2}
\eeq
where
\be
\vec{b}_1=\left({2\pi\over L},-{2\pi\tau_1\over L\tau_2}\right),\;\;
\vec{b}_2=\left(0,{2\pi\over L\tau_2}\right).
\ee

To project the wave function into the LLL, we note that the terms in Eq.~\ref{CFFS1} have factors of the form $e^{i\vec{k_n}\cdot\vec{r}_m} e^{z_m^2-|z_m|^2\over 4\ell^2}$. 
We first write $e^{i\vec{k_n}\cdot\vec{r}_m}=e^{\frac{i}{2}(k_n\bar{z}_m+\bar{k}_nz_m)}$ where $k_n = k_{n,x}+ik_{n,y}$.  The ``direct" projection~\cite{Dev92}
is accomplished by bringing $\bar{z}_m$ to the left and making the replacement $\bar{z}_m\rightarrow 2\ell^2\partial_{z_m}$ (with the understanding that the derivative does not act on $e^{-|z_m|^2/4\ell^2}$)~\cite{Girvin84b,Jain89}. That gives
\begin{widetext}
\be
\label{direct projection}
P_{\rm LLL}e^{i\vec{k}_n\cdot\vec{r}_m} e^{z_m^2-|z_m|^2\over 4\ell^2}=  
e^{-|z_m|^2\over 4\ell^2}     
e^{\frac{i}{2}k_n 2\ell^2\partial_{z_m}}e^{z_m^2\over 4\ell^2}
 e^{\frac{i}{2}\bar{k}_nz_m}
 \equiv
e^{z_m^2-|z_m|^2\over 4\ell^2}\hat{F}_{k_n} (z_m)
\ee
where
\be
\hat{F}_{k_n}(z_m) =e^{-\frac{k_n\ell^2}{4}\left(k_n+2\bar{k}_n\right)}e^{\frac{\mi}{2}(\bar{k}_n+k_n)z_m}e^{i k_n\ell^2\partial_{z_m}}
\label{eqnhatf}
\ee
With this, the CFFS wave function at $\nu=1/2$ can be written as 
\be
\Psi^{\rm CFFS}_{1/ 2,k_{\rm CM}} =e^{\sum_m{z_m^2-|z_m|^2\over 4\ell^2}}\elliptic{{\phi_1\over 4\pi}+{N_\phi-2\over 4}}{-{\phi_\tau\over 2\pi}+N-1}{{2(Z+\sum_j i\ell^2k_j)\over L}}{2\tau}{\rm Det}\left[\hat{F}_{k_n}(z_m)\right]\prod_{i=1}^N J_i
\label{CFFS10}
\ee
\end{widetext}
where
\be
J_i= \prod_{j\neq i} \vartheta\begin{bmatrix}{1\over 2} \\ {1\over 2}\end{bmatrix}\Bigg( {z_i-z_j\over L}\Bigg | \tau \Bigg ).
\ee
Since the Slater determinant is made up of the operators $\hat{F}_{k_n}(z_m)$, this form is impractical for the treatment of large systems. To overcome this issue, we resort to the JK projection \cite{Jain97,Jain97b}. The idea of JK projection is to make in Eq.~\ref{CFFS10}, the replacement 
\be
{\rm Det}\left[\hat{F}_{k_n}(z_m)\right]\prod_{i=1}^N J_i\rightarrow {\rm Det}\left[\hat{F}_{k_n}(z_m)J_m\right]
\ee
and then project each element 
$\hat{F}_{k_n}(z_m)J_m$ separately into the LLL. However, a modification is necessary in $\hat{F}_{k_n}(z_m)J_m$ in the torus geometry to preserve the quasiperiodic boundary conditions: the last factor in Eq.~\ref{eqnhatf} must be replaced by $e^{i 2k_n\ell^2\partial_{z_m}}$.  The final form of the projected CFFS wave function is \cite{Rezayi00,Shao15,Geraedts18,Wang19,Pu17,Pu18,Pu20}:

\begin{widetext}
\beq
\label{CFFS}
P_{\rm LLL} \Psi_{{1\over 2}}^{\rm CFFS}=  e^{\sum_i {z_i^2-\abs{z_i}^2 \over 4 \ell^2}} \Bigg[  \vartheta
\begin{bmatrix}
{\phi_1\over 4\pi}
  + {N_{\phi}-2 \over 4}\\
-{\phi_{\tau}\over 2\pi } + {(N-1)}
\end{bmatrix}
\Bigg({2\left(Z+i\ell^2\sum_j k_j\right) \over L} \Bigg |2 \tau \Bigg)\Bigg] {\rm Det}[G_{k_n}(z_m)] 
\eeq
\beq
G_{k_n}(z_m)= e^{\frac{k_n \ell^2}{4}(k_n+2\bar{k}_n)}e^{\frac{i}{2}(\bar{k}_n+k_n)z_m}\prod_{j,j\neq m }  \vartheta 
\begin{bmatrix}
\frac{1}{2} \\ \frac{1}{2}
\end{bmatrix}
\Bigg(\frac{z_m+\mi \textcolor{red}{2}k_n\ell^2-z_j}{L}|\tau\Bigg)
\eeq
\end{widetext}
Here, we have a nontrivial factor of $2$ for the original translation by $\mi k_n\ell^2$ in $G_{k_n}(z_m)$ to preserve the PBC.

{\it The MR wave function}: The MR wave function, introduced in Ref.~\cite{Moore91}, refers to a special form of CF pairing. In the disk geometry, it is given by
\be
\Psi_{\rm MR}=\exp[-\sum_j|z_j|^2/4\ell^2]{\rm Pf}\left( {1\over z_j-z_k} \right) \prod_{j<k}(z_j-z_k)^2. \label{Pfdisk}
\ee
Here Pf represents Pfaffian, which is defined, for an $N\times N$ antisymmetric matrix $M_{ij}$ (with even $N$), as
\beq
{\rm Pf}\{M_{ij}\}=\frac{1}{2^{N/2}(N/2)!}\sum_{\sigma}\prod_{i=1}^{N/2}M_{\sigma(2i-1)\sigma(2i)}
\eeq
where $\sigma$ labels all permutations. The term ${\rm Pf}\left({1\over z_j-z_k }\right)$ represents a paired state with pair wave function $1/(z_j-z_k)$, and the factor $\prod_{j<k}(z_j-z_k)^2$ converts electrons into composite fermions.

On a torus, the MR wave function can be written as \cite{Greiter92a,Chung07,Read96}
\begin{widetext}
\begin{align} \label{MRPf}
\Psi_{\rm MR}^{(a,b)} &= e^{\sum_i \frac{z_i^2-|z_i|^2}{4\ell^2}}\Bigg\{\vartheta
\begin{bmatrix}
{\phi_1\over 4\pi}
 + {N_{\phi}-2 \over 4} +{(1-2a)\over 4}\\
-{\phi_{\tau}\over 2\pi } - {(N_{\phi}-2)\over 2} -{(1-2b)\over 2}
\end{bmatrix}
\Bigg({2Z \over L} \Bigg |2 \tau \Bigg) \Bigg \} {\rm Pf}\Bigg ( { \vartheta \begin{bmatrix}
a \\ b
\end{bmatrix}({z_i -  z_j \over L}|\tau) \over \vartheta\begin{bmatrix}
\frac{1}{2} \\ \frac{1}{2}
\end{bmatrix}
({z_i -  z_j \over L}|\tau) } \Bigg )  \Bigg[\prod_{i<j} \vartheta 
\begin{bmatrix}
\frac{1}{2} \\ \frac{1}{2}
\end{bmatrix}
\Bigg(\frac{z_i - z_j}{L}|\tau \Bigg)\Bigg]^2 
\end{align}
\end{widetext}
The parameters $(a,b)$ take values $(0,1/2)$, $(1/2,0)$ or $(0,0)$. The resulting wave functions correspond to the three-fold topological degeneracy of MR states and lie in different Haldane pseudo-momentum (i.e. relative momentum) sectors \cite{Haldane85b} $(N/2,0)$, $(0,N/2)$, $(N/2,N/2)$, respectively. These states are exactly degenerate for the 3-body interaction for which the MR wave function is the exact ground state and are believed to become degenerate for the Coulomb interaction in the thermodynamic limit. The correspondence of the MR wave function in Eq.~\ref{MRPf} with the more familiar disk geometry form of Eq.~\ref{Pfdisk} can be seen from the facts that the Pfaffian part in Eq.~\ref{MRPf} is analogous to ${\rm Pf}{1\over (z_j-z_k)}$ since $\vartheta\begin{bmatrix}
\frac{1}{2} \\ \frac{1}{2}
\end{bmatrix}
({z_i -  z_j \over L}|\tau) $ vanishes for $z_i=z_j$ while
$\vartheta \begin{bmatrix}
a \\ b
\end{bmatrix}({z_i -  z_j \over L}|\tau)$ does not vanish for $z_i=z_j$ for the chosen values of $(a,b)$.

\section{Construction of BCS wave function for composite fermions}

The MR wave function in Eq.~\ref{MRPf} represents a topological $p_x-ip_y$ pairing of composite fermions and has exotic properties such as excitations with non-abelian braiding statistics. However, unlike the BCS wave function, it does not have any variational parameters, and it is unclear how the MR state may be connected to the CFFS by tuning the pairing strength. In this section, we construct a BCS wave function for composite fermions with two variational degrees of freedom. This wave function explicitly reduces to the CFFS in one limit. We calculate its energy as a function of the variational parameters and find that, to a good approximation, the minimum energy wave function also has the highest overlap with the exact SLL ground state. We take this as theoretical evidence for pairing in the $5/2$ state.

First consider a system of spin-polarized electrons with an attractive interaction. The ground state of the system can be described by the BCS wave function. The mean field BCS Hamiltonian can be written as \cite{Read99}
\begin{equation}
    H = \sum_{\vec{k}} \epsilon_{\vec{k}} c_{\vec{k}}^{\dagger}c_{\vec{k}} + \sum_{\vec{k}} ({1\over 2}{\Delta_{\vec{k}}} c_{\vec{k}}^{\dagger}c^{\dagger}_{-\vec{k}} + h.c.)
\end{equation}
where $h.c.$ means hermitian conjugation,  $\epsilon_{\vec{k}} = \hbar ^2 |\vec{k}|^2/2m-\hbar ^2 \abs{\vec{k}_F}^2/2m$ and  the gap function $\Delta_{\vec{k}}=\Delta \; (k_x -ik_y)$ is taken to have $p$-wave symmetry. Here $m$ is the electron mass and $\vec{k}_F$ is the Fermi wave vector. 
We take $\Delta$ to be a real number. The BCS wave function for electrons can be written as \cite{Read99}\cite{DeGennes99} 
\begin{equation}
\ket{\Psi_{\rm BCS}} = \prod'_{\vec{k}}(u_{\vec{k}} + v_{\vec{k}}c_{\vec{k}}^{\dagger}c_{-\vec{k}}^{\dagger})\ket{0}
\end{equation} 
where $\prod'$ means that each $\vec{k},-\vec{k}$ pair only appears once. [We choose $\phi_1=\phi_2=\pi$ so that the reciprocal lattice vectors are given by $\vec{k} = (n_1+1/2)\vec{b_1} + (n_1+1/2) \vec{b_2}$;  with this choice, $\vec{k}=0$ is absent and the allowed wave vectors appear in pairs $\pm \vec{k}$.]
The state $\ket{0}$ is the null state with no electrons.
Following the standard Bogoliubov transformation, we can obtain the relation
\begin{equation}
\label{gk1}
g_{\vec{k}}\equiv\frac{v_{\vec{k}}}{u_{\vec{k}}}= \frac{\epsilon_{\vec{k}} - \sqrt{\epsilon_{\vec{k}}^2 +|\Delta_{\vec{k}}|^2}}{\Delta_{\vec{k}}^*}=-g_{-\vec{k}}
\end{equation}
The particle number is not conserved in $\ket{\Psi_{\rm BCS}}$. After projecting $\ket{\Psi_{\rm BCS}}$ into a fixed particle-number sector, the real space BCS wave function is given by \cite{DeGennes99}
\be
\label{BCS electron}
\Psi_{\rm BCS}(\vec{r}_1,....,\vec{r}_{N}) = {\rm Pf}(g(\vec{r}_i-\vec{r}_j))
\ee
where the function $g$ is antisymmetric under exchange of two particles. In the plane-wave basis, we have 
\be
g(\vec{r}_i-\vec{r}_j)=\sum_{\Vec{k}_n} g_{\Vec{k}_n} e^{i\Vec{k}_n\cdot\left(\Vec{r_i}-\Vec{r}_j\right)}.
\ee

We now write the BCS wave function for composite fermions at $\nu=1/2$ as 
\begin{equation}
\label{CF-BCS}
    \Psi _{\frac{1}{2}}^{\rm BCS} = P_{\rm LLL} {\rm Pf}\left(\sum _{\vec{k}_n} g_{\vec{k}_n} e^{i\vec{k}_n\cdot\left(\vec{r}_i-\vec{r}_j\right)}\right) \Psi^{\rm L}_{1/2}
\end{equation}
where $\Psi^{\rm L}_{1/2}$ is given in Eq.~\ref{Laughlin}.
We assume that the form of $g_{\vec{k}}$ is still given by Eq.~\ref{gk1}, but with the electron mass $m$ replaced by the CF mass $m^*$. 

We need to project the wave function into the LLL. 
For ``direct" projection, we can follow the approach described for the CFFS to project the paired plane wave $e^{i\vec{k}_n\cdot \left(\vec{r}_1-\vec{r}_2\right)}$:
\begin{widetext}
\ba
     P_{\rm LLL} e^{i\vec{k}_n\cdot \left(\vec{r}_1-\vec{r}_2\right)} e^{\frac{z_1^2+z_2^2 - \abs{z_1}^2-\abs{z_2}^2}{4\ell^2}} &=&  e^{\frac{z_1^2+z_2^2 - \abs{z_1}^2-\abs{z_2}^2}{4\ell^2}}e^{-\frac{k_n\ell^2}{2}(k_n+2\Bar{k}_n)}e^{\frac{i}{2}(z_1 - z_2)(k_n+\Bar{k}_n)}e^{ik_n\ell^2 \partial _{z_1}} e^{-ik_n\ell^2 \partial _{z_2}}  \nonumber\\
&\equiv&e^{\frac{z_1^2+z_2^2 - \abs{z_1}^2-\abs{z_2}^2}{4\ell^2}}\hat{F}_n(z_1,z_2)
\ea

The CF-BCS wave function can then be written as 
\be
 \Psi_{\frac{1}{2}}^{\rm BCS} = e^{\sum_i \frac{z_i^2 - |z_i|^2}{4\ell^2}}\elliptic{{\phi_1\over 4\pi}+{N_\phi-2\over 4}}{-{\phi_\tau\over 2\pi}+N-1}{{2Z\over L_1}}{2\tau}{\rm Pf}\left[\sum_ng_{\vec{k}_n}\hat{F}_n(z_i,z_j)\right]\prod_iJ_i
\ee
At this point, each matrix element $\sum_ng_{\vec{k}_n}\hat{F}_n(z_i,z_j)$ in the Pfaffian is an operator. One can directly confirm that the operators inside the Pfaffain commute with the center-of-mass part of $\Psi^{\rm L}_{1/2}$, since the momentum $\vec{k}_n$ is always paired with $-\vec{k}_n$. 
\end{widetext}

The above form is not amenable to calculations for systems with more than eight particles. For that reason, we appeal to the 
JK projection. As discussed previously, the JK projection in its simplest version fails to conserve the periodic boundary conditions, and it is necessary to modify it. The modification for the Jain states and the CFFS was discussed in Refs. ~\cite{Pu17,Pu18}. For the CF-BCS wave function, the situation is even more complicated because we need to bring both $J_i$ and $J_j$ into the Pfaffian, to write it as
\be
{\rm Pf}\left[\sum_ng_{\vec{k}_n}\hat{F}_n(z_i,z_j)\right]\prod_iJ_i\rightarrow 
{\rm Pf}\left[\sum_ng_{\vec{k}_n}\hat{F}_n(z_i,z_j)J_i J_j\right],
\ee
and then project each matrix element separately. In order to preserve the boundary conditions, we 
find that we need to replace the $e^{ik_n\ell^2 \partial _{z_i}} e^{-ik_n\ell^2 \partial _{z_j}} $ factor in $\hat{F}_n(z_i,z_j)$ by $e^{ik_n\ell^2 \hat{D}^{(j)} _{z_i}} e^{-ik_n\ell^2 \hat{D}^{(i)} _{z_j}}$, where the new derivative operator $\hat{D}^{(j)}_{z_i}$ is defined as
\begin{widetext}
\be
\hat{D}^{(j)}_{z_i}\elliptic{1/2}{1/2}{z_i-z_l\over L}{\tau}\equiv
\begin{cases}
{\partial\over \partial{z_i}}\elliptic{1/2}{1/2}{z_i-z_l\over L}{\tau} \quad {\rm if}\quad l=j\\
2{\partial\over \partial{z_i}}\elliptic{1/2}{1/2}{z_i-z_l\over L}{\tau} \quad {\rm if}\quad l\neq j
\end{cases}
\ee

The final form for the JK projected wave function is:

\begin{equation}
\label{BCS_paired}
    \Psi_{\frac{1}{2}}^{\rm BCS} = e^{\sum_i \frac{z_i^2 - |z_i|^2}{4\ell^2}}\Bigg\{\vartheta
\begin{bmatrix}
{\phi_1\over 4\pi}
 + {N_{\phi}-2 \over 4}\\ 
-{\phi_{\tau}\over 2\pi } + {N-1}
\end{bmatrix}
\Bigg({2Z \over L} \Bigg |2 \tau \Bigg) \Bigg \} {\rm Pf}(M_{ij})
\end{equation}
in which the Pfaffian matrix element is:
\beq
\label{LLL pfa}
    M_{ij} = \sum_{k_n}g_{\vec{k}_n}e^{-\frac{\ell^2}{2}k_n(k_n+2\Bar{k}_n)}e^{\frac{i}{2}(z_i-z_j)(k_n+\Bar{k}_n)}\Bigg(\vartheta  \begin{bmatrix}
{1 \over 2} \\ {1 \over 2 }
\end{bmatrix}\Bigg(\frac{z_i +ik_n\ell^2 - (z_j-ik_n\ell^2)}{L}|\tau \Bigg)\Bigg)^2 \nonumber \\ \Bigg \{ \prod_{\substack{r \\r \neq i,j}}   \vartheta  \begin{bmatrix}
{1 \over 2} \\ {1 \over 2 }
\end{bmatrix}\Bigg(\frac{z_i + i2k_n\ell^2- z_r}{L}|\tau \Bigg) \prod_{\substack{m \\m \neq i,j}}  \vartheta  \begin{bmatrix}
{1 \over 2} \\ {1 \over 2 }
\end{bmatrix}\Bigg(\frac{z_j - i2k_n\ell^2- z_m}{L}|\tau \Bigg) \Bigg \}
\eeq

$M_{ij}$ is odd under exchange of $i$ and $j$, as can be seen by noting that $M_{ij} = \sum_{k}M_{ij}^{k_n}$, and that $M_{ij}^{k_n}$ is odd under $\vec{k}_n\rightarrow -\vec{k}_n$ and exchange of $i$ and $j$:
\begin{align}
M_{ij}^{k_n} &= g_{\vec{k}_n}e^{-\frac{\ell^2}{2}k_n(k_n+2\Bar{k}_n)}e^{\frac{i}{2}(z_i-z_j)(k_n+\Bar{k}_n)}\Bigg(\vartheta  \begin{bmatrix}
{1 \over 2} \\ {1 \over 2 }
\end{bmatrix}\Bigg(\frac{z_i +ik_n\ell^2 - (z_j-ik_n\ell^2)}{L}|\tau \Bigg)\Bigg)^2 \Bigg \{ \cdots \Bigg \} \nonumber
\\&
=-g_{-\vec{k}_n}e^{-\frac{\ell^2}{2}(-k_n)((-k_n)-2\Bar{k}_n)}e^{\frac{i}{2}(z_j-z_i)(-k_n-\Bar{k}_n)}\Bigg(\vartheta  \begin{bmatrix}
{1 \over 2} \\ {1 \over 2 }
\end{bmatrix}\Bigg(\frac{z_j -ik_n\ell^2 - (z_i+ik_n\ell^2)}{L}|\tau \Bigg)\Bigg)^2
\Bigg \{ \cdots \Bigg \} = -M_{ji}^{-k_n}
\end{align}
where $\{ \cdots \}$ contains the last two terms in Eq.~\ref{LLL pfa}. 
The sum over the $(\vec{k}_n,-\vec{k}_n)$ pairs give us $M_{ij}=-M_{ji}$ for $i \neq j$.
The boundary conditions can be verified as shown in the appendix \ref{appx-PBC}.  

\begin{figure*}
\begin{minipage}[b]{0.3\linewidth}
\includegraphics[width=\linewidth]{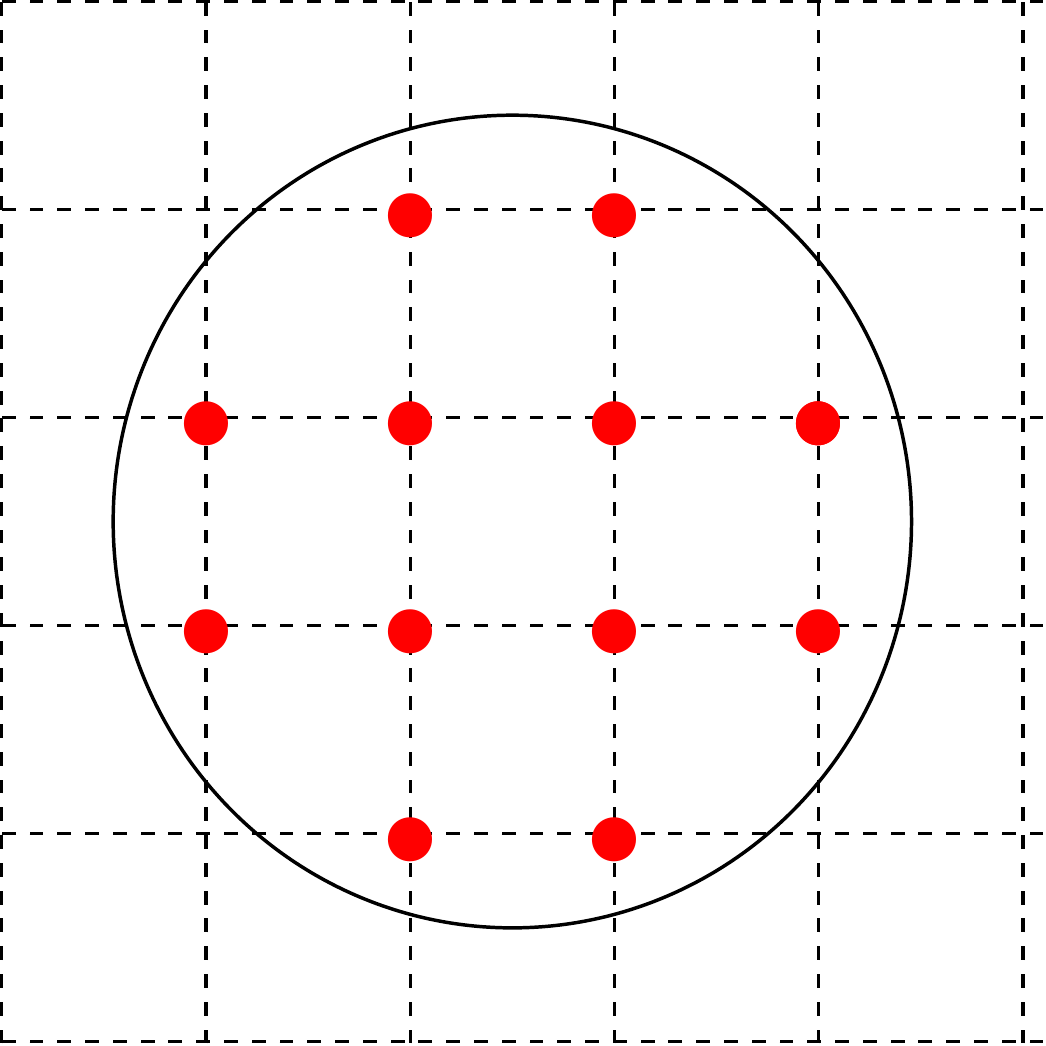} 
\end{minipage}
\quad
\begin{minipage}[b]{0.3\linewidth}
\includegraphics[width=\linewidth]{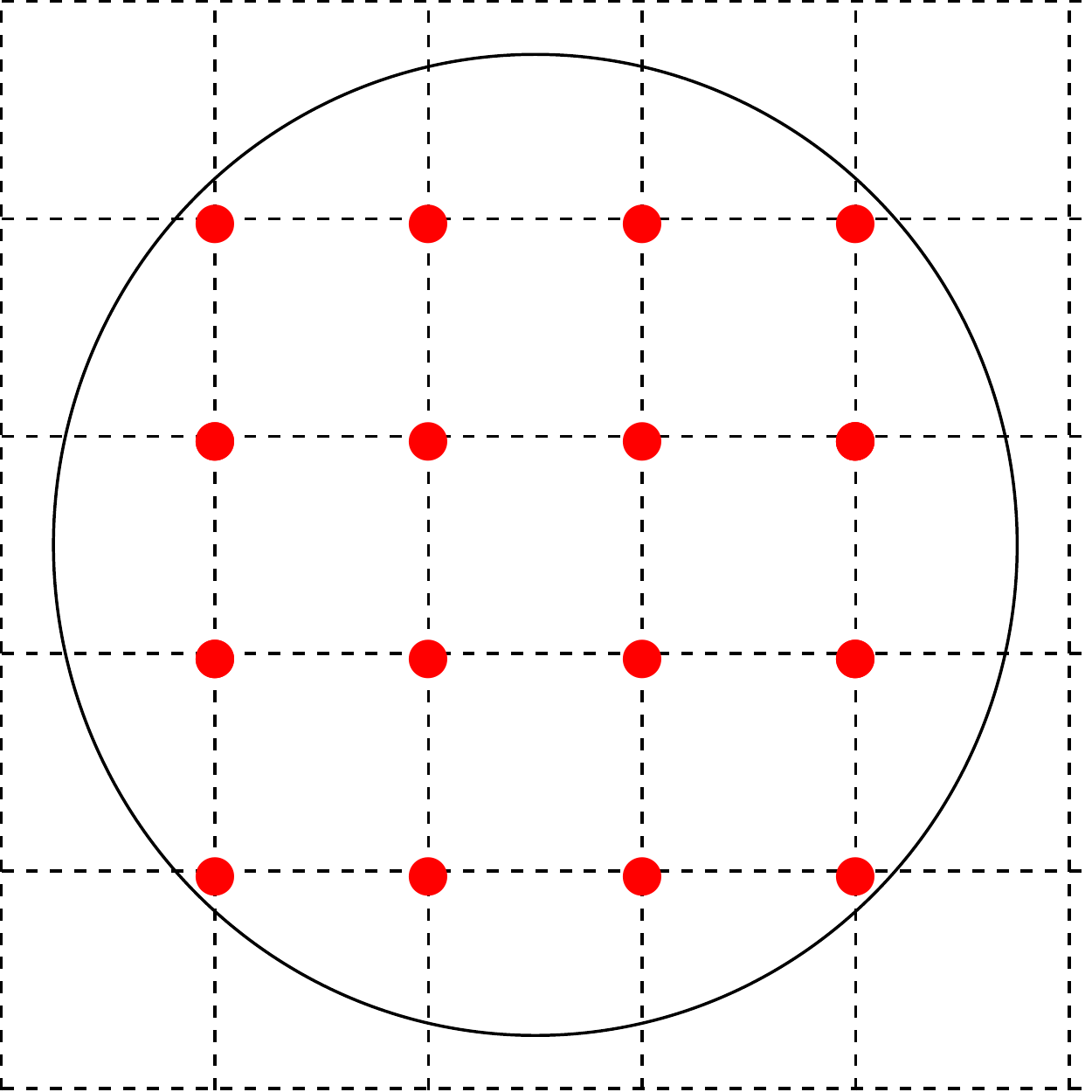} 
\end{minipage}
\quad
\begin{minipage}[b]{0.3\linewidth}
\includegraphics[width=\linewidth]{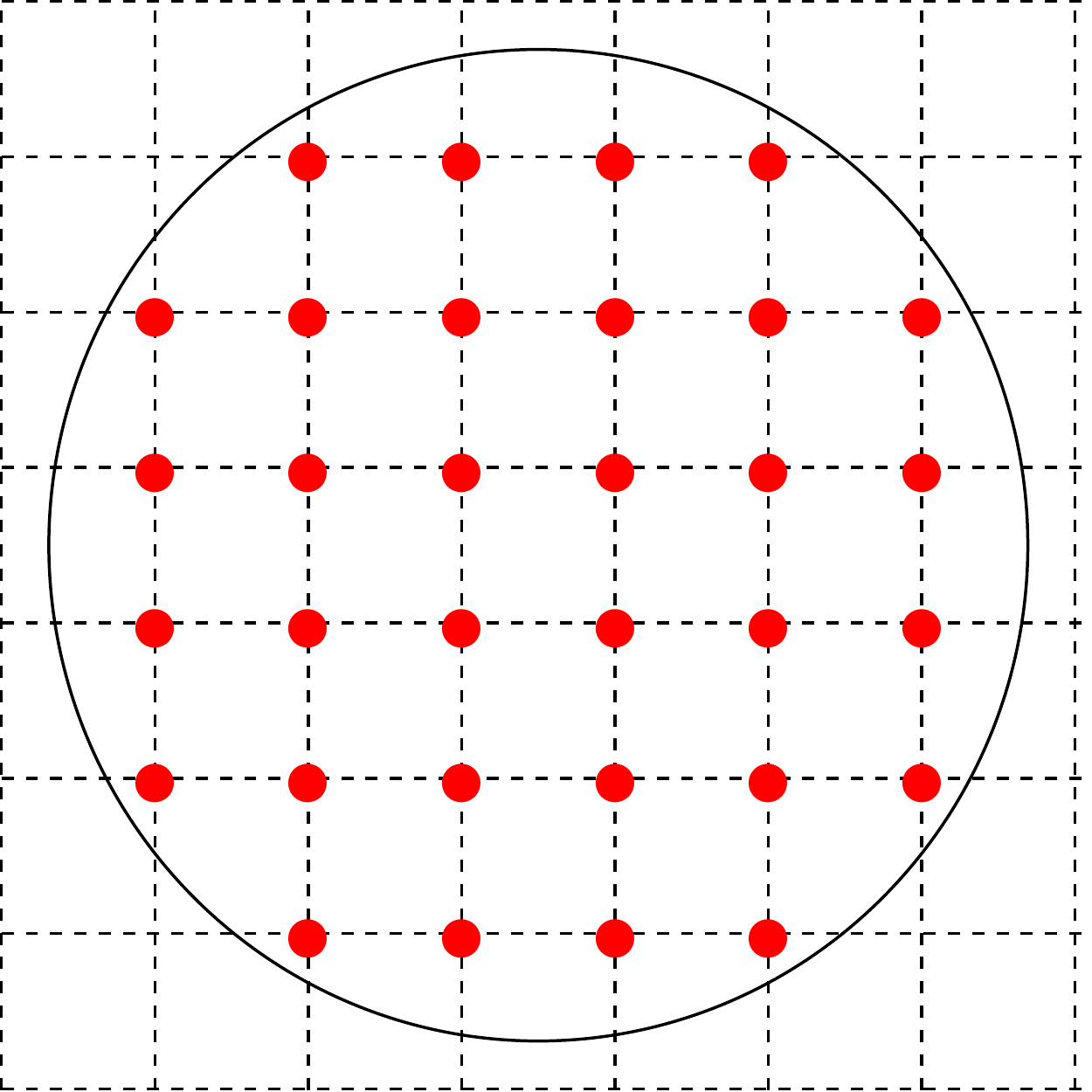} 
\end{minipage}
\caption{\label{configuration}The configurations of Fermi sea used in our calculations for $N=12$ (left panel), $N=16$ (middle panel) and $N=32$ (right panel). We also show the approximate circular Fermi surface.
}
\end{figure*}

To parameterize our BCS-paired wave function, we define a dimensionless variational parameter $\delta_{k_F}$, which we refer to as the gap parameter,  
\beq 
\delta_{k_F}={\Delta|k_F| \over \hbar^2|k_F|^2/2m^*}
\eeq
The information of the pairing strength $\Delta$ and effective mass $m^*$ are both encoded in the definition of $\delta_{k_F}$. Now Eq.~\ref{gk1} can be re-written as:
\beq
\label{gk2}
g_{\vec{k}_n} ={ {\hbar^2\abs{k_n}^2\over 2m^*}-{\hbar^2\abs{k_F}^2\over 2m^*}-\sqrt{\Bigg({\hbar^2\abs{k_n}^2\over 2m^*}-{\hbar^2\abs{k_F}^2\over 2m*}\Bigg)^2+\abs{\Delta k_n}^2} \over \Delta (k_{n,x}+ik_{n,y})} ={\abs{k_n}^2 - \abs{k_F}^2-\sqrt{(\abs{k_n}^2 - \abs{k_F}^2)^2+\delta_{k_F}^2|k_F|^2\abs{k_n}^2}\over \delta_{k_F}(k_{n,x}+ik_{n,y})|k_F|}
\eeq
\end{widetext}

We introduce another variational parameter $k_{\rm cutoff}$, which is a cutoff on $\vec{k}_n$ in $\sum_{\vec{k}_n}$ in Eq.~\ref{BCS_paired}. More specifically, we set that $|g_{\vec{k}_n}|=0$ for $|\vec{k}_n|>k_{\rm cutoff}$.	This may be viewed as being analogous to the Debye cutoff for the gap in the BCS theory. We will minimize the energy of the wave function with respect to these two parameters $\delta_{k_F}$ and $k_{\rm cutoff}$.

We have shown above that our BCS wave function satisfies the correct boundary conditions. Another important property of a physical wave function in the torus geometry is modular covariance. As mentioned above, the geometry of torus is parameterized by the modular parameter $\tau$. The correspondence between $\tau$ and the geometry is not one-to-one. The geometry is unchanged under the modular transformations $\mathcal{T}:\tau\rightarrow \tau+1$, $\mathcal{S}:\tau\rightarrow -1/\tau$ and any combination of these two transformations \cite{Gunning62}. Obviously, any physical observables should be invariant under these transformations. This requires that the mixing of a degenerate set of wave functions is closed under these transformations. The wave functions that satisfy this property are said to be modular covariant \cite{Fremling14,Fremling19,Pu20}. The modified JK projection for Jain states and CFFSs have been shown to be modular covariant \cite{Fremling19,Pu20,Pu20b}. In Appendix~\ref{appx-mod}, we show the CF-BCS-paired wave function in Eq.~\ref{CF-BCS} before LLL projection is modular covariant, and its ``direct" projection into the LLL also produces a modular covariant wave function. This Appendix also shows that the JK projection of the CF-BCS wave function in Eq.~\ref{BCS_paired} does {\it not} produce a modular covariant wave function. This makes the situation problematic because it is this form of the wave function that allows calculations for large systems. Fortunately, we find that the JK projected wave function provides an energy that is very close to that of the direct projected wave function. This provides justification for using the JK projected wave functions for our variational study below.

\section{Numerical study of pairing instability}

Having constructed the wave function, we proceed, as in the BCS theory, to find the value of the parameters, $\delta_{k_F}$ and the momentum cutoff $k_{\rm cutoff}$, that minimize the energy. We wish to do this  for both $\nu=1/2$ and $\nu=5/2$, to capture the remarkably different physics at these two filling fractions. For $\nu=1/2$ we simply work with (a periodic version of) the Coulomb interaction. For $\nu=5/2$, we use an effective interaction in the LLL to mimic the SLL coulomb interaction by matching their Haldane pseudopotential coefficients. This was earlier done in Ref.~\cite{Park98b}, which showed that an accurate effective interaction is:
\beq
\label{Veff}
V^{\rm eff}(r) = \frac{e^2}{\epsilon} \Big\{\frac{1}{r}+a_1e^{-\alpha_1 r^2} + a_2r^2e^{-\alpha_2 r^2} \Big\}
\eeq
The best-fitted parameters are $a_1 = 117.429,a_2=-755.468 $, $\alpha_1=1.3177$, and $\alpha_2=2.9026$, which guarantee that the first four pseudopotential coefficients are the same as the second LL Coulomb pseudopotentials. While calculating the energy on torus geometry, the k-space summation of the interaction should be used~ \cite{Yoshioka83}. The details of the numerical calculations are given in Appendix \ref{appx-int}. We neglect corrections due to finite thickness and LL mixing throughout this work.

\begin{figure}[t]
\includegraphics[width=\linewidth]{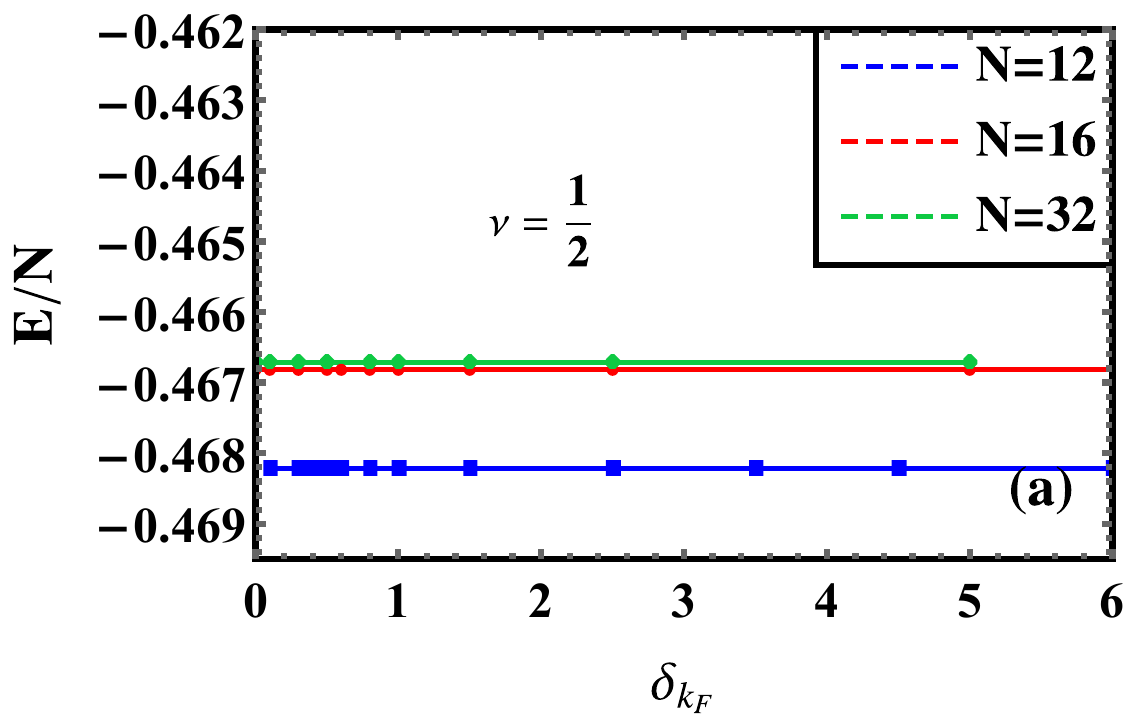} 
\includegraphics[width=\linewidth]{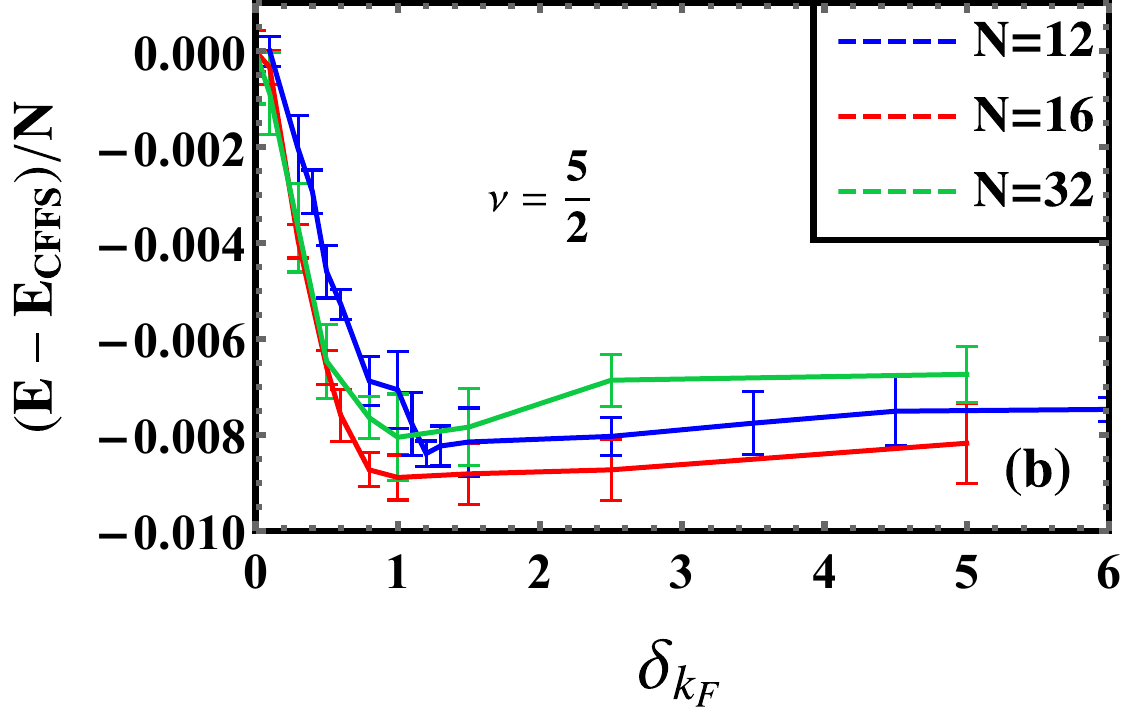} 
\caption{\label{energy-plot} The Coulomb energy per particle for (a) $\nu=1/2$, (b) $\nu=5/2$ as a function of the parameter $\delta_{k_F}$ for different system sizes. For each value of $\delta_{k_F}$, minimum energy is obtained by varying the momentum cutoff. The energies are quoted 
in units of $e^2/\epsilon\ell$; for  $\nu=5/2$, the energies are plotted relative to the CFFS energy. 
At $\nu=1/2$ the CFFS has the lowest energy for all $\delta_{k_F}$ (for $\nu=1/2$, the error bars have been omitted, which are on the order of 0.00001).  
}
\end{figure}

We have performed our calculation for systems with 12, 16 and 32 particles, because these produce fairly circular Fermi seas for even $N$. The approximate magnitude of $k_F$ is estimated using the relation:
\beq
\pi |k_F|^2=N |\vec{b_1} \cross \vec{b_2}|.
\eeq 
In Fig.~\ref{configuration}, we show the $k$-space configurations of CFFS for these systems, 
with the solid black lines showing the approximate Fermi surfaces. For both $\nu=1/2$ and $\nu=5/2$, we find the minimum energy by considering a range of values for $\delta_{k_F}$ and minimizing the energy for each $\delta_{k_F}$ by varying $k_{\rm cutoff}$.

{The energies per particle are shown in  Fig.~\ref{energy-plot} for both $\nu=1/2$ and $\nu=5/2$ as a function of $\delta_{k_F}$. This illustrates the most notable finding of our work: at $\nu=5/2$, }
the energy minimum for SLL occurs at $\delta_{k_F}\approx 1.2$, {indicating the presence of CF pairing.} In contrast, the minimum energy at $\nu=1/2$ is obtained for $\delta_{k_F}=0$, i.e. for the CFFS, which is consistent with an absence of pairing. However, we note that due to the discreteness of the momentum lattice, our work does not rule out, strictly speaking, a very weak pairing at $\nu=1/2$.

\begin{figure}[t]
\centering
\includegraphics[width=\linewidth]{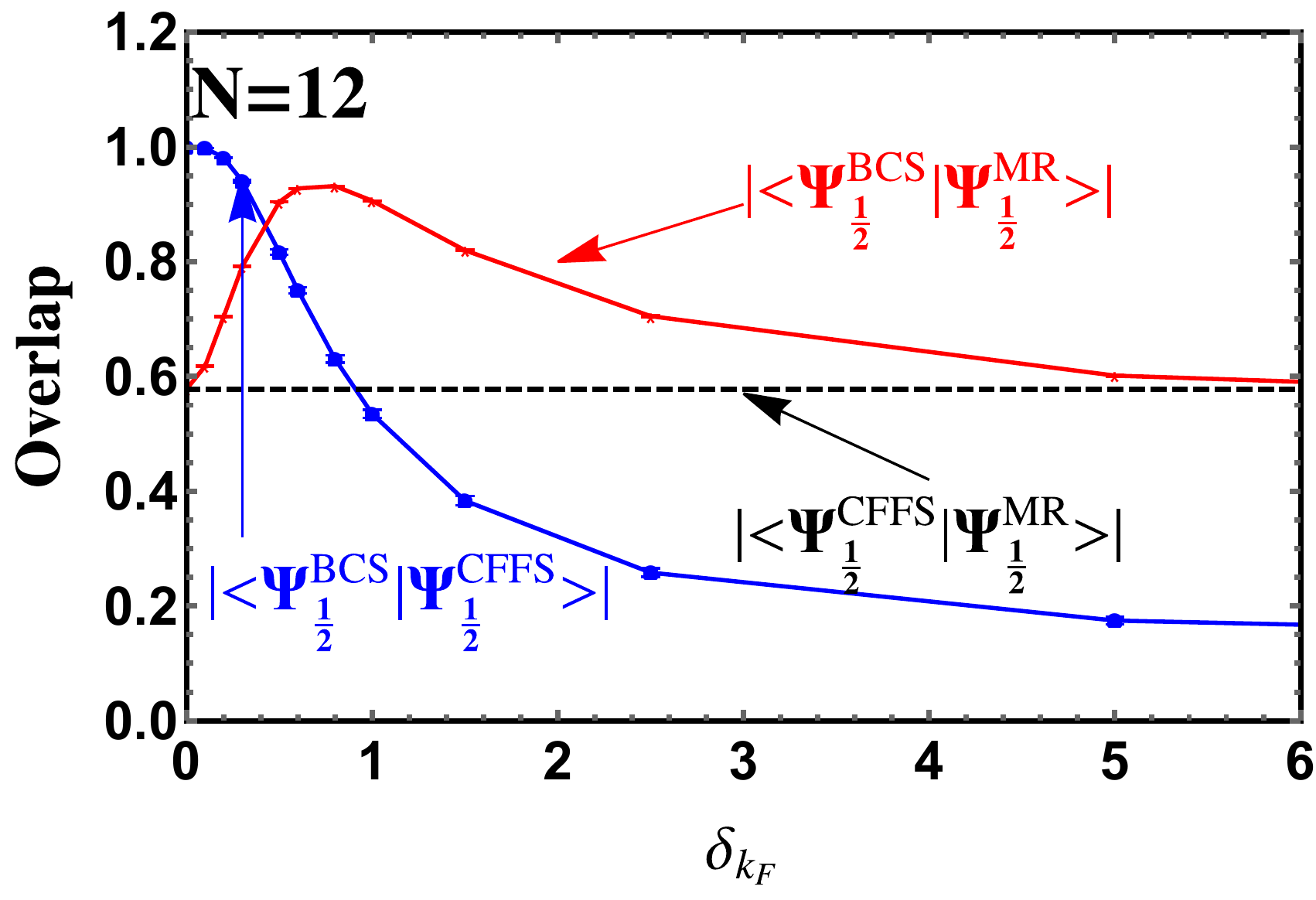} 
\includegraphics[width=\linewidth]{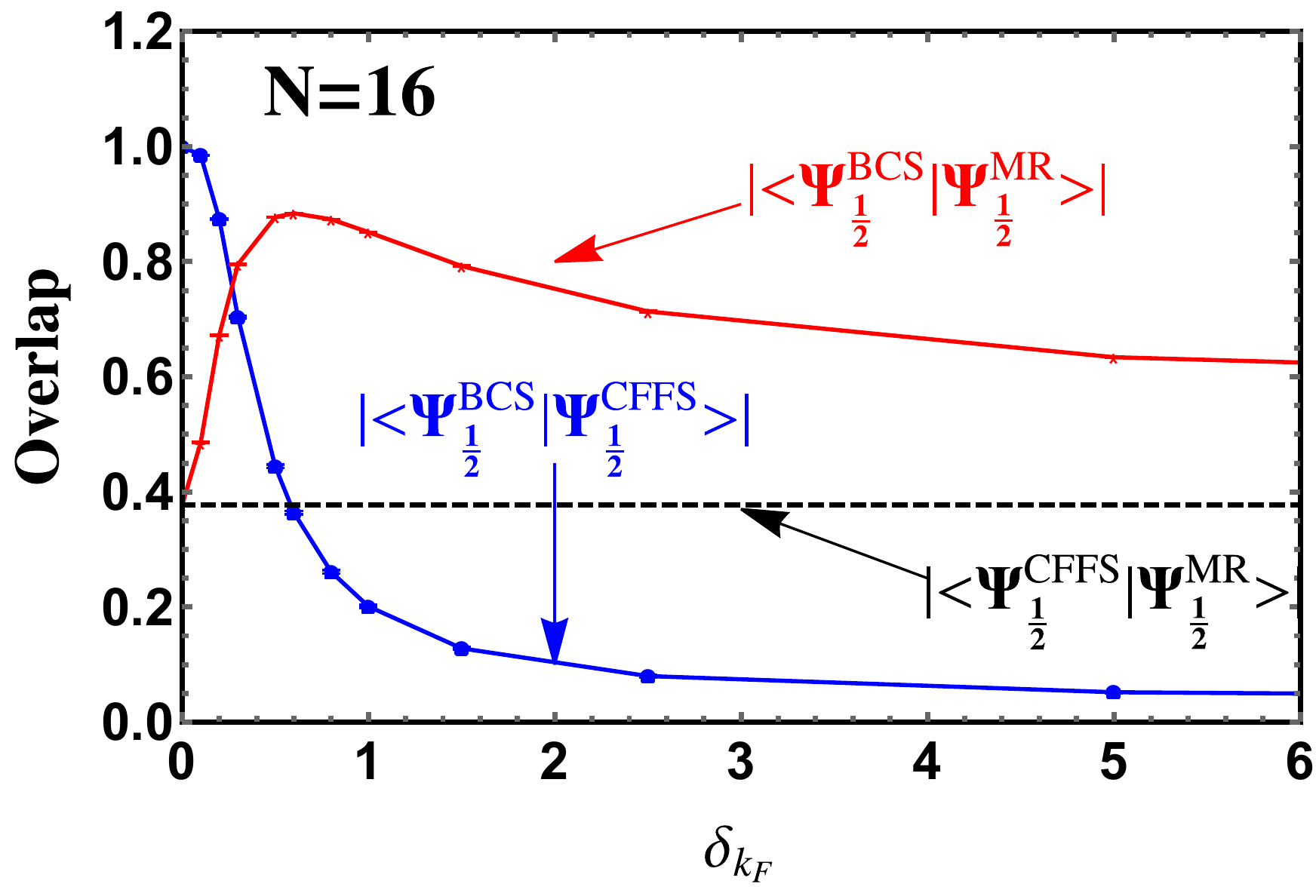} 
\caption{\label{overlap}The overlaps between the CF-BCS wave function, MR wave function, and the CFFS wave function as a function of $\delta_{k_F}$ for systems with $N=12$ (upper panel) and $N=16$ (lower panel). 
}
\end{figure}

\begin{figure}
\begin{center}
\includegraphics[width=0.9\columnwidth]{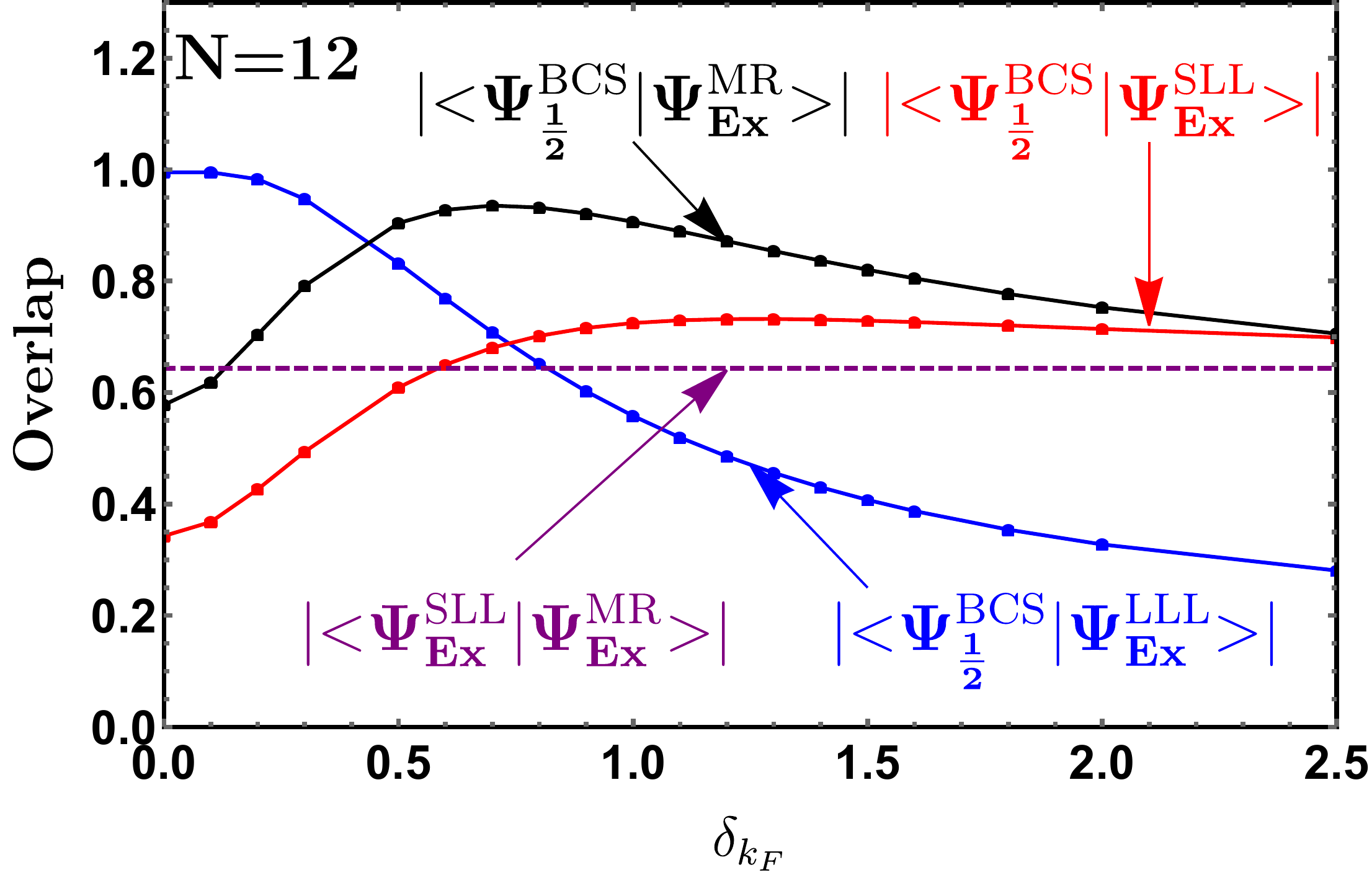} 
\end{center}
\caption{\label{exact} Comparisons with the exact ground state for $N=12$ particles. $\ket{\Psi ^{\rm BCS}}$, $\ket{\Psi ^{\rm LLL}_{\rm Ex}}$, $\ket{\Psi ^{\rm SLL}_{\rm Ex}}$, $\ket{\Psi ^{\rm MR}_{\rm Ex}}$ refer to the CF-BCS wave function, the exact LLL ground state, the exact SLL ground state and the MR state, 
respectively. The overlaps of the CF-BCS wave function with $\ket{\Psi ^{\rm LLL}_{\rm Ex}}$, $\ket{\Psi ^{\rm SLL}_{\rm Ex}}$ and $\ket{\Psi ^{\rm MR}_{\rm Ex}}$ are shown as a function of $\delta_{k_F}$ for a 12 particle system.  The purple line marks the overlap between MR state and SLL state: $|\bra{\Psi ^{\rm MR}_{\rm Ex}}\ket{\Psi ^{\rm SLL}_{\rm Ex}}| = 0.64358$. The maximum overlap between CF-BCS and SLL ground state is reached around $\delta_{k_F}\approx 1.3$, with a value 0.73203.  For reference, we have $|\bra{\Psi ^{\rm LLL}_{\rm Ex}}\ket{\Psi ^{\rm SLL}_{\rm Ex}}| = 0.34365$ and  $|\bra{\Psi ^{\rm LLL}_{\rm Ex}}\ket{\Psi ^{\rm MR}_{\rm Ex}}| = 0.60022$. The method for calculating the above overlaps has been outlined in Appendix \ref{ex-overlap}.
}
\end{figure}

To ascertain how the CF BCS wave function compares with the MR wave function, we compute the
overlap of the CF BCS wave function with the MR wave function 
for different values of the variational parameter $\delta_{k_F}$ for $N=12,16$ particles. The overlaps are shown in Fig.~\ref{overlap}, which also displays the overlap of the CF-BCS state with the CFFS. (The overlaps are calculated for wave functions within the same Haldane pseudo-momentum sector $(k_1,k_2)$.) The overlaps between different trial wave functions, as shown in Fig.~\ref{overlap}, are obtained using the Monte Carlo algorithm. When $\delta_{k_F}\rightarrow 0$, the overlap between the CF BCS wave function and CFFS is $1$, as expected. 
The CF BCS wave function has the highest overlap of $\sim 0.94$ ($\sim 0.88$) with the MR state for $\delta_{k_F}\approx 0.7$ (Fig.~\ref{exact}) for $N=12$ ($N=16$) particles. 
We also obtain the overlap of the CF-BCS state with the exact LLL, SLL and MR state as shown in Fig.~\ref{exact}. The method used to obtain the overlaps with the exact states is outlined in appendix \ref{ex-overlap}. The overlap of the CF BCS state with the exact LLL state decreases as $\delta_{k_F}$ increases. The CF BCS state has the highest overlap with the SLL ground state at around $\delta_{k_F}\approx 1.3$, which is close to the lowest energy state in SLL. 
We find that the lowest energy CF BCS wave function (MR wave function) has an overlap $0.73$ ($0.64$) with the SLL Coulomb ground state.

\begin{figure}[t]
\begin{center}
\includegraphics[width=0.9\columnwidth]{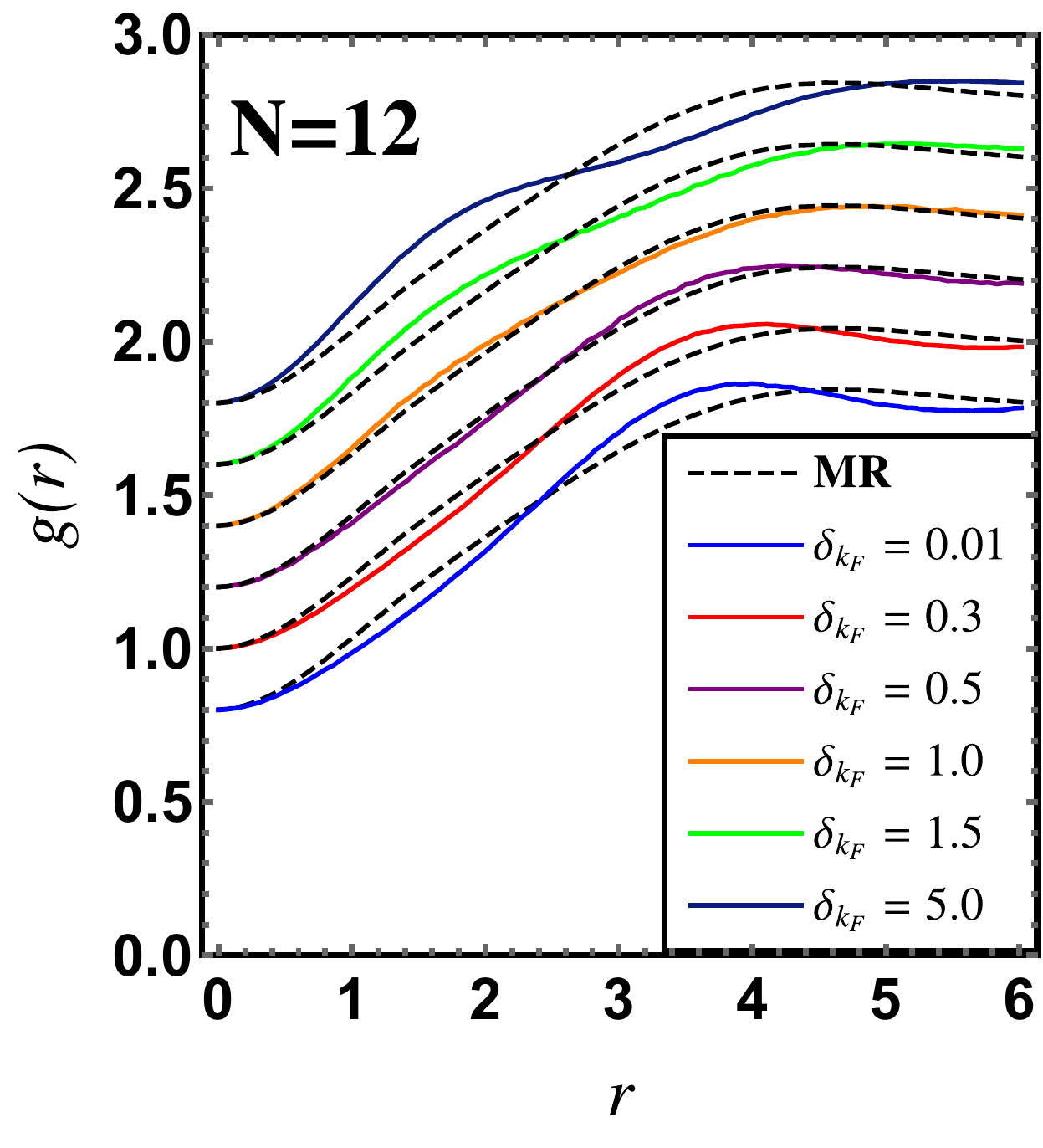} 
\end{center}
\caption{\label{pair-corr}Plot of $g(r)$, the pair correlation function of the CF-BCS wave function, for the 12 particle system. The distance $r$ is measured in units of the magnetic length $\ell$. The pair-correlation functions for different $\delta_{k_F}$  have been shifted vertically for clarity; for each curve, we have $g(r=0)=0$. The pair correlation function for the MR wave function is shown by the dashed black curve. 
}
\end{figure}

Evidence of the pairing is seen in the pair correlation function plotted in Fig.~\ref{pair-corr}, where a short distance bump develops with increasing $\delta_{k_F}$. For comparison, we also show the pair correlation function for the MR wave function. The pair correlation function of the BCS wave function is in best agreement with the pair correlation function of the MR wave function for $\delta_{k_F}\approx 0.5-1.0$, which is also near the optimal value of $\delta_{k_F}$. Both of these results suggest that, for the 5/2 state, the CF-BCS wave function is close to the MR wave function for certain parameters. 
We mention here, for completeness, that the MR and our CF-BCS wave functions have the same pairing form. For long-distance correlation, which corresponds to the small $k$ limit, the $g_{\vec{k}}$ in Eq.~\ref{gk2} is of the form $1\over k_x+ik_y$, which transforms into the form $1\over z$ in the real space, which is the same as in the MR state \cite{Ma19}.

Within the BCS theory, the physical gap is related to the condensation energy as
\beq
\label{conden}
E(\delta_{k_F} \rightarrow 0)- E(\delta_{min})= \rho(E_F){ \tilde{\Delta}^2 \over 2}
\eeq
where $\delta_{min}$ is the value of $\delta_{k_F}$ that minimizes the energy, and $\tilde{\Delta}$ is the physical gap, and $\rho(E_F)$ is the density of states at the Fermi energy. The gap evaluated in this fashion is given by
\beq
{\tilde{\Delta}\over {e^2/\epsilon \ell}}\sim \sqrt{0.1 {E_F\over e^2/ \epsilon \ell} }
\eeq
 for the systems we have studied. Unfortunately, we are not able to obtain this gap numerically, because we do not have a good estimate for the CF Fermi energy at $\nu=5/2$. It is also unclear to what extent Eq.~\ref{conden} above, which applies to  weakly interacting electrons, is valid for composite fermions.

\section{Spin Singlet pairing}
The BCS paired wave function considered so far is for spin-polarized composite fermions. Theoretical calculations indicate that, in the limit of vanishing Zeeman energy, the ground state  at $\nu=1/2$ is a spin singlet CFFS~\cite{Park98,Balram17}. Thus, it is worth asking whether an instability into a spin-singlet pairing occurs in the limit of vanishing Zeeman energy. The g-factor can be made to vanish in GaAs quantum wells by application of hydrostatic pressure~\cite{Leadley97}; alternatively, one can consider multi-valley systems where the valley index plays the role of spin~\cite{Padmanabhan10,Feldman12,Kott14}.

Following the analogy of spinless fermions, we construct a spin-singlet BCS-paired CF wave function. The starting point is the spin-singlet BCS wave function of electrons 
\beq
\ket{\Psi_{\rm BCS}} = \prod _{\vec{k}} (u_{\vec{k}} + v_{\vec{k}} c_{\vec{k}\uparrow}^{\dagger}c_{-\vec{k}\downarrow}^{\dagger}) \ket{0}
\eeq
After projecting this state to a sector with $2N$ electrons, the real space wave function (without normalization) can be written as \cite{Bouchaud88}
\begin{equation}
\Psi_{\mathrm{\rm BCS}}^{\mathrm{singlet}}(r_1,r_2,..,r_{2N})=\rm{Pf}
\begin{bmatrix}
0 & M^{\uparrow \downarrow} \\
-M^{\uparrow \downarrow \rm{T}} & 0 
\end{bmatrix}
\end{equation}
where $[M^{\uparrow \downarrow}]_{ij}=\sum_{\vec{k}} g_{\vec{k}} e^{i\vec{k}.(\vec{r}_{i\uparrow}-\vec{r}_{j\downarrow})}$ is a symmetric matrix under exchange of indices. For the singlet wave function with $s$-wave pairing, we have
\beq
g_{\vec{k}}= \frac{\epsilon_{\vec{k}} - \sqrt{\epsilon_{\vec{k}}^2 +\Delta^2}}{\Delta},
\eeq
where $\Delta$ is a real number and has dimensions of energy (not to be confused with $\Delta$ for the spin polarized case).
Using the property for a $N \cross N$ dimensional matrix $A$
\beq
\rm{Pf} \begin{bmatrix}
0 & A \\
-A^{\rm{T}} & 0 
\end{bmatrix}
= (-1)^{\textit{N}(\textit{N}-1)/2}\rm{Det}[\textit{A}],
\eeq
it can be shown that the above wave function can be written as \cite{Bajdich08,Bouchaud88}.
\begin{equation}
\Psi_{\mathrm{\rm BCS}}^{\mathrm{singlet}}(r_1,r_2,..,r_{2N})= \rm{Det}[\sum_{\vec{k}} g_{\vec{k}} e^{i\vec{k}.(\vec{r}_{i\uparrow}-\vec{r}_{j\downarrow})}]
\end{equation}
where $i=1,2,...,N$ and $j=N+1,...,2N$ represent the indices for spin-up and spin-down electrons, respectively. The above form of the wave function is convenient for numerical calculations. The composite-fermionized wave function for the singlet CF-BCS state is given by
\begin{equation}
\Psi_{\frac{1}{2}}^{\mathrm{singlet}}= P_{\rm LLL}{\rm{Det}}[\sum_{\vec{k}} g_{\vec{k}} e^{i\vec{k}.(\vec{r}_{i\uparrow}-\vec{r}_{j\downarrow})}]{\Psi}_{1/2}^{\rm L} 
\end{equation}
It can be projected into the LLL in the same fashion as the $p$-wave paired state.

\begin{figure}[t]
\begin{center}
\includegraphics[width=0.9\columnwidth]{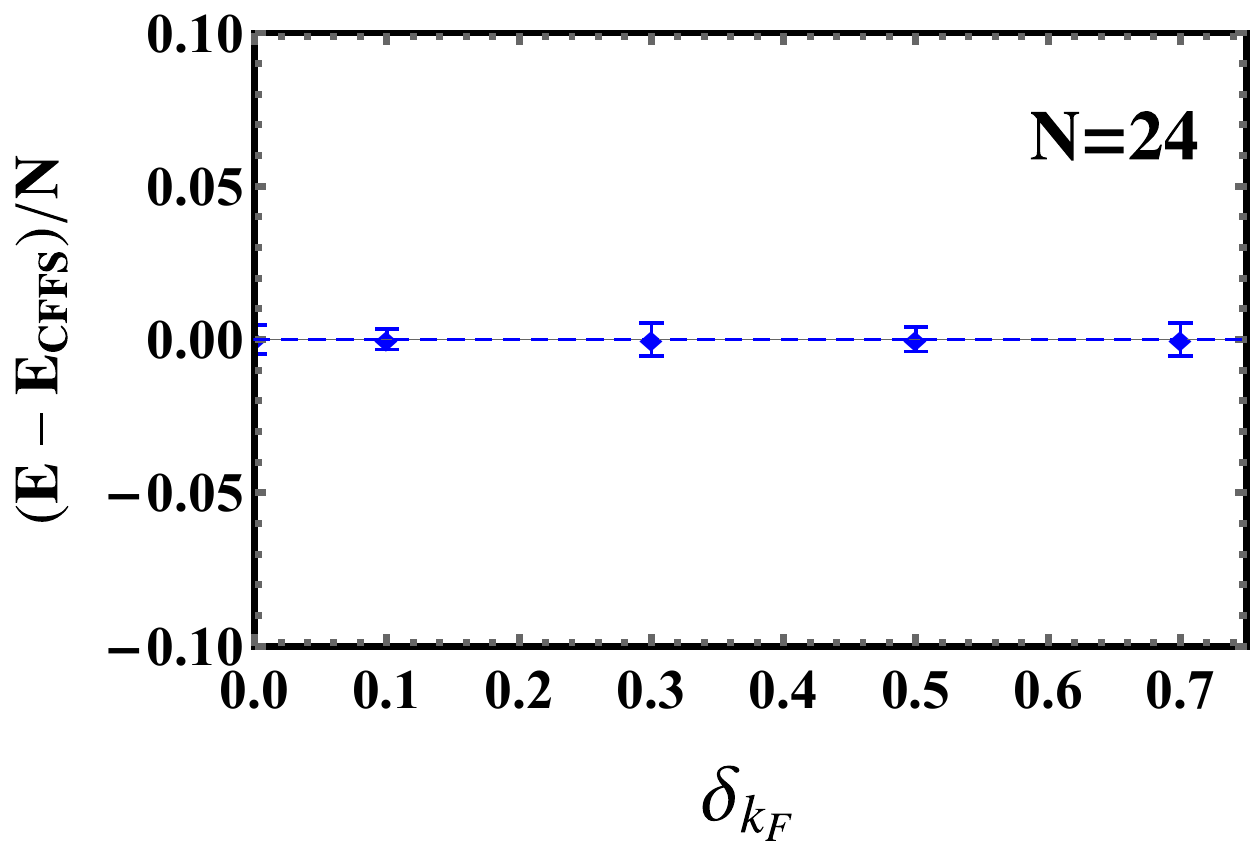} 
\end{center}
\caption{\label{Singlet}The energy per particle for the spin singlet CF-BCS wave function at $\nu=1/2$ as a function of $\delta_{k_F}$. The spin-singlet CFFS is seen to have the lowest energy, indicating an absence of pairing instability. The energies are plotted relative to the CFFS energy in units of $e^2/\epsilon \ell$. 
}
\end{figure}

The LLL energy plot in Fig.~\ref{Singlet} indicates that for all values of $\delta_{k_F}$, 
the minimum energy state is obtained at for $k_{\rm cutoff}=k_{F}$, which corresponds to the spin singlet CFFS. There is thus no spin-singlet pairing instability.

\section{Hall viscosity of paired BCS wave function}

One of the topological quantities of a fractional quantum Hall state is its Hall viscosity $\eta^A$ \cite{Avron95}. As proposed in Ref.~\cite{Read09}, $\eta^A$ is related to the orbital spin \cite{Wen92}, or the ``shift" $\sh=N/\nu-N_\phi$ in the spherical geometry, as:
\be
\label{hall visc}
\eta^A=\sh{\hbar \over 4}{\rho}.
\ee
Here $\rho$ is the 2D density.
Hall viscosity serves to distinguish different topological states that have the same Hall conductance. Eq.~\ref{hall visc} has been derived for the Laughlin state, Pfaffian state, and Jain states by various approaches \cite{Read09,Tokatly09,Read11,Lapa18,Lapa18b,Cho14,Pu20}. It has also been numerically confirmed for the Laughlin and Pfaffian states in Ref.~\cite{Read11} and for the Jain states in Refs.~\cite{Fremling14,Pu20}.

The Hall viscosity is computed through the Berry curvature in the $\tau$ space, which captures the adiabatic change of gapped state with the shear deformation of the torus:  
\beq
\label{berry curv}
\eta^{A} = -{\hbar \tau_{2}^2 \over V} \mathcal{F}_{\tau_1,\tau_2}
\eeq
where $V$ is the total area and $\mathcal{F}_{\tau_1,\tau_2}$ is the berry curvature in $\tau$ space. The berry curvature is defined as
\beq
\mathcal{F}_{\tau_1,\tau_2}=-2\mathbf{Im}\bra{\partial\Psi \over \partial \tau_1}\ket{\partial\Psi \over \partial \tau_2}
\eeq
\begin{figure}[t]
\centering
\includegraphics[width=\linewidth]{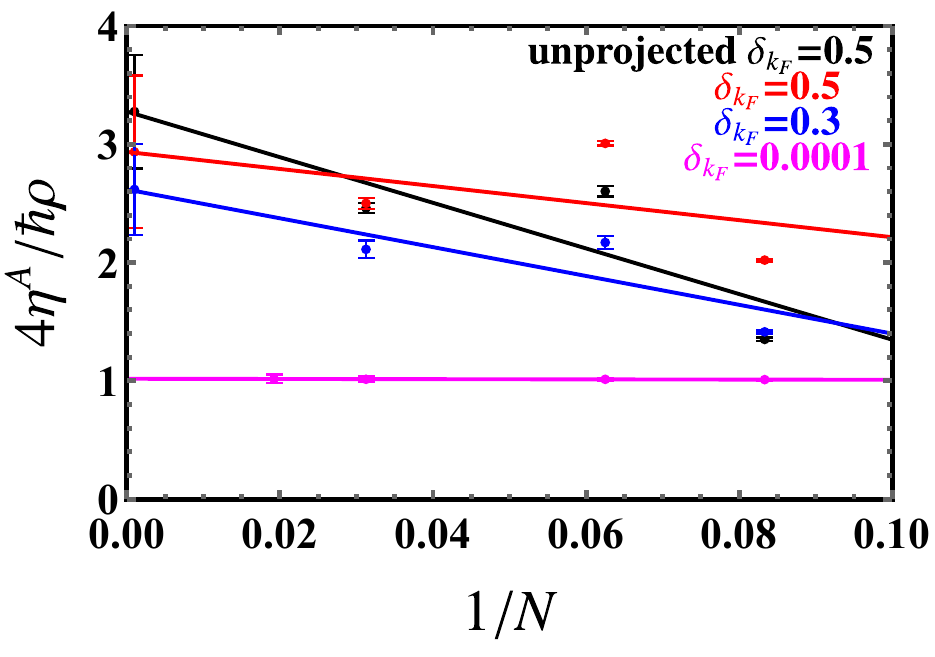} 

\caption{\label{fig-Hall-Pf}The Hall viscosity of the JK projected CF-BCS wave function for different values of $\delta_{k_F}$ as a function of $1/N$. The Hall viscosity for the unprojected CF-BCS wave function is also shown for  $\delta_{k_F}=0.5$.  The $k_{\rm cutoff}$ for each $\delta_{k_F}$ corresponds to minimum energy. The magenta points for $\delta_{k_F}=0.0001$ essentially correspond to the CFFS.
}
\end{figure}

We have calculated the Hall viscosity $\eta^A$ of the CF-BCS state employing Eq.~\ref{berry curv}, which we evaluate using the Monte Carlo method. The evaluation of the Pfaffian for large system sizes was accomplished following the method described in Ref.~\cite{Wimmer12}.  Our results are shown in Fig.~\ref{fig-Hall-Pf}.  In the limit $\delta_{k_F} \rightarrow 0$, which is the CF-Fermi sea, we obtain the same Hall viscosity as that reported in Ref.~\cite{Pu20b}. When $\delta_{k_F}$ increases, the Hall viscosity changes. As shown in Fig.~\ref{fig-Hall-Pf}, the Hall viscosity shows strong finite size fluctuations, but in the thermodynamic limit, {it is consistent with} ${3\hbar \over 4}{\rho}$, the Hall viscosity of the MR state. {It is noted that this value is expected for any non-zero value of the gap parameter $\delta_{k_{F}}$.}

{A caveat is in order here.  As discussed in Appendix \ref{appx-mod}, the JK projected CF-BCS wave function does not satisfy modular covariance, in contrast to the ``direct" projected CF-BCS wave function. However, as shown in Fig.~\ref{Direct-results} in Appendix \ref{appx-mod},  the energy of the JK-projected wave function is close to that of the ``direct" projected wave function for $N=4$ particles for different values of the modular parameter $\tau$. We can expect that the JK-projected BCS wave function gives at least a good first  approximation to the Hall viscosity. We also obtained the Hall viscosity for the unprojected CF-BCS wave function, which is modular covariant. As shown in Fig.~\ref{fig-Hall-Pf}, the Hall viscosities of the JK projected and the unprojected CF-BCS wave function are consistent with that of the MR state in the thermodynamic limit. Although the values of the Hall viscosity for the JK projected and the unprojected CF-BCS wave functions are different for finite systems, the variational tendencies with respect to the system size are similar.}

\section{Conclusions}

We have constructed a $p$-wave paired BCS wave function for composite fermions on the torus with two variational parameters. We have shown how the JK projection can be modified to satisfy the periodic boundary conditions. Our CF-BCS wave function enables us to calculate energy, overlap, pair correlation, and Hall viscosity for fairly large system sizes. We find a pairing instability for $\nu= {5\over 2}$ but not for $\nu={1\over 2}$. Furthermore, we find that the parameters that produce minimum energy at $\nu=5/2$ also maximize the overlap with the exact Coulomb ground state. These results overall support the notion of CF pairing mechanism at $\nu=5/2$.

We note that our study does not include the effect of LL mixing, and thus is not capable of distinguishing between the MR Pfaffian and the anti-Pfaffian wave functions.
Our results also suggest absence of $s$-wave spin-singlet pairing of composite fermions at $\nu={1 \over 2}$ in the limit of zero Zeeman energy. As a future direction, it would be interesting to investigate pairing of composite fermions in other contexts~\cite{Bonesteel96,Scarola02b,Moran12,Mukherjee12,Kim19,Faugno19,Li19}.

\section{Acknowledgments}

The work was supported by the U. S. Department of Energy, Office of Basic Energy Sciences, under Grant no. DE-SC0005042. We acknowledge Advanced CyberInfrastructure computational resources provided by The Institute for CyberScience at The Pennsylvania State University. We thank Ajit C. Balram for useful discussions and M. Wimmer for the open-source PFAPACK library, used for the numerical evaluation of the Pfaffian of matrices. Some of the numerical diagonalizations were performed using the DiagHam package, for which we are grateful to its authors.

\begin{appendix}

\section{Boundary conditions}
\label{appx-PBC}
In this Appendix, we show that the wave function in Eq.\ref{BCS_paired} satisfies the correct periodic boundary conditions.

For the wave function with $2N$ particles, the Pfaffian of the matrix $M$ will have terms like $M_{12}M_{34}...M_{pj}...$, product of $N$ different elements of $M$. Each index in the subscript occurs only once, and there can be permutations in the ordering of indices. The translation of a single matrix element along the $\tau$ direction gives us, for $p\neq i,j$:
\begin{widetext}
\ba
  &&  T_p(L\tau) M_{ij} \\\nonumber
&=&  
    \Bigg \{ \sum_{k}g_{k}e^{-\frac{\ell^2}{2}k(k+2\Bar{k})}e^{\frac{i}{2}(z_i-z_j)(k+\Bar{k})} e^{i\pi (\frac{2(z_i +i2k\ell^2 - z_p)}{L}-\tau)} e^{i\pi (\frac{2(z_j -i2k\ell^2 - z_p)}{L}-\tau)} \prod_{\substack{r \\r \neq i,j}}  \vartheta \begin{bmatrix} {1 \over 2} \\ {1 \over 2} \end{bmatrix}\Bigg(\frac{z_i + i2k\ell^2- z_r}{L}|\tau \Bigg)\\\nonumber
   && \prod_{\substack{m \\m \neq i,j}}  \vartheta \begin{bmatrix} {1 \over 2} \\ {1 \over 2} \end{bmatrix}\Bigg(\frac{z_j - i2k\ell^2- z_m}{L}|\tau \Bigg)   \Bigg(\vartheta \begin{bmatrix} {1 \over 2} \\ {1 \over 2} \end{bmatrix}\Bigg(\frac{z_i +i2k\ell^2 - z_j}{L}|\tau \Bigg)\Bigg)^2  \Bigg \}\\ \nonumber
    &=& e^{i2\pi \frac{(z_i+z_j)}{L}}e^{-i\frac{4\pi z_p}{L}}e^{-i2\pi \tau} M_{ij} 
\ea
and for $p=i$ or $p=j$, we get
\ba
&& T_p(L \tau) M_{pj}\\\nonumber
 &=& T_p(L\tau) \Bigg \{ \sum_{k}g_{k}e^{-\frac{\ell^2}{2}k(k+2\Bar{k})}e^{\frac{i}{2}(z_p-z_j)(k+\Bar{k})}  \prod_{\substack{r \\r \neq p,j}}  \vartheta \begin{bmatrix} {1 \over 2} \\ {1 \over 2} \end{bmatrix}\Bigg(\frac{z_p + i2k\ell^2- z_r}{L}|\tau \Bigg) \prod_{\substack{m \\m \neq p,j}}  \vartheta \begin{bmatrix} {1 \over 2} \\ {1 \over 2} \end{bmatrix}\Bigg(\frac{z_j - i2k\ell^2- z_m}{L}|\tau \Bigg) \\\nonumber
 &&\Bigg(\vartheta \begin{bmatrix} {1 \over 2} \\ {1 \over 2} \end{bmatrix}\Bigg(\frac{z_p +i2k\ell^2 - z_j}{L}|\tau \Bigg)\Bigg)^2  \Bigg \} \\\nonumber
    &=&  \sum_{k}g_{k}e^{-\frac{\ell^2}{2}k(k+2\Bar{k})}e^{\frac{i}{2}(z_p-z_j)(k+\Bar{k})} e^{\frac{i}{2}L\tau(k+\Bar{k})} e^{-i (2N-2)\pi( \frac{2(z_p+i2k\ell^2)}{L}+ \tau)}e^{i2\pi\frac{\sum_a' z_a}{L}} e^{-i2\pi(\frac{2(z_p+i2k\ell^2-z_j)}{L}+ \tau)}\\\nonumber
    &&\prod_{\substack{r \\r \neq p,j}}  \vartheta \begin{bmatrix} {1 \over 2} \\ {1 \over 2} \end{bmatrix}\Bigg(\frac{z_p + i2k\ell^2- z_r}{L}|\tau \Bigg)  \prod_{\substack{m \\m \neq p,j}}  \vartheta \begin{bmatrix} {1 \over 2} \\ {1 \over 2} \end{bmatrix}\Bigg(\frac{z_j - i2k\ell^2- z_m}{L}|\tau \Bigg)  \Bigg(\vartheta \begin{bmatrix} {1 \over 2} \\ {1 \over 2} \end{bmatrix}\Bigg(\frac{z_p +i2k\ell^2 - z_j}{L}|\tau \Bigg)\Bigg)^2 \\\nonumber
    &=& \sum_{k}g_{k}e^{-\frac{l^2}{2}k(k+2\Bar{k})}e^{\frac{i}{2}(z_p-z_j)(k+\Bar{k})} e^{\frac{i}{2}L\tau(k+\Bar{k})} e^{-i 2N\pi( \frac{2z_p}{L}+\frac{i4kl^2}{L}+ \tau)}e^{i2\pi\frac{\sum_a' z_a}{L}} e^{i2\pi(\frac{2z_j}{L})}\\\nonumber
     &&\prod_{\substack{r \\r \neq p,j}}  \vartheta \begin{bmatrix} {1 \over 2} \\ {1 \over 2} \end{bmatrix}\Bigg(\frac{z_p + i2kl^2- z_r}{L}|\tau \Bigg)  \prod_{\substack{m \\m \neq p,j}}  \vartheta \begin{bmatrix} {1 \over 2} \\ {1 \over 2} \end{bmatrix}\Bigg(\frac{z_j - i2kl^2- z_m}{L}|\tau \Bigg)  \Bigg(\vartheta \begin{bmatrix} {1 \over 2} \\ {1 \over 2} \end{bmatrix}\Bigg(\frac{z_p +i2kl^2 - z_j}{L}|\tau \Bigg)\Bigg)^2 \\\nonumber
     &=& e^{-i \frac{4N\pi z_p}{L}} e^{-i2N\pi \tau} e^{i2\pi\frac{\sum_a' z_a}{L}} e^{i2\pi(\frac{2z_j}{L})} M_{pj}
\ea
where $\sum _a '= \sum_{\substack{ \\a\neq p,j}} $.
Combining the above two results, we obtain
\begin{align*}
    T_p(L\tau) M_{12}M_{34}...M_{pj}... =& \Bigg[ \Bigg\{e^{i2\pi \frac{(z_1+z_2)}{L}}e^{-i\frac{4\pi z_p}{L}}e^{-i2\pi \tau} \Bigg\} \Bigg\{e^{i2\pi \frac{(z_3+z_4)}{L}}e^{-i\frac{4\pi z_p}{L}}e^{-i2\pi \tau} \Bigg\}...
    \\&
    \Bigg \{ e^{-i \frac{4N\pi z_p}{L}} e^{-i2N\pi \tau} e^{i2\pi\frac{\sum_a' z_a}{L}}e^{i2\pi(\frac{2z_j}{L})} \Bigg \}... \Bigg]  M_{12}M_{34}...M_{pj}...
    \\=& e^{i4\pi \frac{\sum_{a}'z_a}{L}}e^{-i(N-1)4\pi \frac{z_p}{L}}  e^{-i \frac{4N\pi z_p}{L}} e^{-i2N\pi \tau}e^{-i2(N-1)\pi \tau} e^{i2\pi(\frac{2z_j}{L})} M_{12}M_{34}...M_{pj}...
    \\= & e^{i4\pi \frac{Z}{L}} e^{-i \frac{8N\pi z_p}{L}}e^{-i4N\pi \tau}e^{i2\pi \tau} M_{12}M_{34}...M_{pj}...
\end{align*}
These equations imply that the phase factor from each term in the expansion of the Pfaffian is independent of permutation the indices. The COM part satisfies the relation 
\be
T_p(L\tau)\Bigg\{\vartheta
\begin{bmatrix}
{\phi_1\over 4\pi}
 + {N_{\phi}-2 \over 4}\\ 
-{\phi_{\tau}\over 2\pi } + {N-1}
\end{bmatrix}
\Bigg({2Z \over L} \Bigg |2 \tau \Bigg) \Bigg \}  = e^{i\phi _{\tau}} e^{-i2\pi \tau} e^{-i\frac{4\pi Z}{L}}  \Bigg\{\vartheta
\begin{bmatrix}
{\phi_1\over 4\pi}
 + {N_{\phi}-2 \over 4}\\ 
-{\phi_{\tau}\over 2\pi } + {N-1}
\end{bmatrix}
\Bigg({2Z \over L} \Bigg |2 \tau \Bigg) \Bigg \} 
\ee 
Putting this all together, we finally have 
 
\begin{equation}
    T_p(L\tau)\Bigg\{\vartheta
\begin{bmatrix}
{\phi_1\over 4\pi}
 + {N_{\phi}-2 \over 4}\\ 
-{\phi_{\tau}\over 2\pi } + {N-1}
\end{bmatrix}
\Bigg({2Z \over L} \Bigg |2 \tau \Bigg) \Bigg \} {\rm Pf}(M_{ij}) = e^{i(\phi _{\tau}-N_{\phi}\pi (\frac{2z_p}{L} + \tau))}\Bigg\{\vartheta
\begin{bmatrix}
{\phi_1\over 4\pi}
 + {N_{\phi}-2 \over 4}\\ 
-{\phi_{\tau}\over 2\pi } + {N-1}
\end{bmatrix}
\Bigg({2Z \over L} \Bigg |2 \tau \Bigg) \Bigg \} {\rm Pf}(M_{ij})
\end{equation}
which is exactly what the periodic boundary condition requires. In the other direction, the periodic boundary condition is satisfied in a similar way.

\end{widetext}

\section{Interaction Energy}
\label{appx-int}
On a torus, the interaction is periodic i.e.
\beq
V(\vec{r}+ m\vec{L} + n\vec{L}\tau) = V(\vec{r})
\eeq
where $m$ and $n$ are integers. The periodic form  for the Coulomb interaction on torus is given by:
\beq
\label{LLL-coulomb}
V_C(r) = {1 \over L^2\rm{Im}({\tau})}\sum_{\vec{q}}{2\pi\over q} e^{i\vec{q} \cdot \vec{r}}
\eeq
\beq
\vec{q} = \left( {2\pi m \over L}, -{2\pi \tau _1 m \over L \tau _2} +{2\pi  n \over L \tau _2} \right)
\eeq
where $2\pi/q$ is the Fourier transformed form of the $1/r$ term. 
For our energy calculations, we have used a rectangular torus. 

To calculate the SLL energies, we  need the Fourier transform of $V_{\rm eff}$ in Eq.~\ref{Veff}.
The first term is treated as above. The Fourier transforms of the other two terms in the effective interaction are:
\beq
\int e^{-\alpha r^2} e^{-i\vec{q}\cdot\vec{r}}d^2\vec{r} =\left({\pi \over \alpha}\right) e^{-q^2\over 4 \alpha}
\eeq

\beq
\int \vec{r}^2 e^{-\alpha r^2} e^{-i\vec{q}\cdot\vec{r}}d^2\vec{r} = \left({\pi \over 2\alpha^2}\right) e^{-q^2\over 4 \alpha}\left(2-{q^2 \over 2 \alpha}\right)
\eeq
The effective interaction can thus be written as 
\beq
V_{\rm eff}={1 \over L^2\rm{Im}(\tau)}\sum_{\vec{q}}V(q) e^{i\vec{q} \cdot \vec{r}}
\eeq
with 
\beq
V(q) = {2\pi\over q} + a_1\left({\pi \over \alpha_1}\right) e^{-q^2\over 4 \alpha _1} + a_2\left({\pi \over 2 \alpha^2_2}\right) e^{-q^2\over 4 \alpha _2}\left(2-{q^2 \over 2 \alpha _2}\right) \nonumber
\eeq

In the above form, we have not included the interaction of a particle in the principal region with its periodic images. The self-interaction energy for the LLL (i.e. for the Coulomb interaction) is given by \cite{Yoshioka83,Bonsall77} :
\beq
W & = & -{e^2\over \epsilon \sqrt{L^2\abs{\tau}}} \left[ 2 - \sum _{mn}' \varphi_{-{1 \over 2}}(\pi(|\tau| m^2 + |\tau| ^{-1}n^2))\right] \nonumber \\
\varphi _{n}&=&\int _1^{\infty} dt e^{-zt}t^n 
\eeq
The prime on the summation indicates that the term $m=0,n=0$ is excluded. The final expression for the energy per particle can be written as 
\beq
E = W + \frac{1}{N}\sum _{i<j}V(\vec{r_i-\vec{r_j}})
\label{tot_energy}
\eeq
The LLL energy can be obtained by plugging Eq.~\ref{LLL-coulomb} into the above equation. In the LLL, $\vec{q}=0$ term in the summation in Eq.~\ref{LLL-coulomb} is excluded since it gets canceled by the electron-background and background-background interactions. A cutoff of $\abs{m},\abs{n} \leq 20 $ in Eq.~\ref{LLL-coulomb} is sufficient to obtain the energy \cite{Pu17}. 

For the SLL, we do not have an explicit expression for the self-interaction energy. However, this does not pose any difficulty since we are interested in the change of the energy rather than its absolute value. For a given system size with the same boundary conditions, the self-interaction does not vary with $\delta_{k_F}$.

\section{Modular covariance of the CF-BCS wave function}
\label{appx-mod}

\begin{figure*}
\begin{minipage}[b]{0.4\linewidth}
\includegraphics[width=\linewidth]{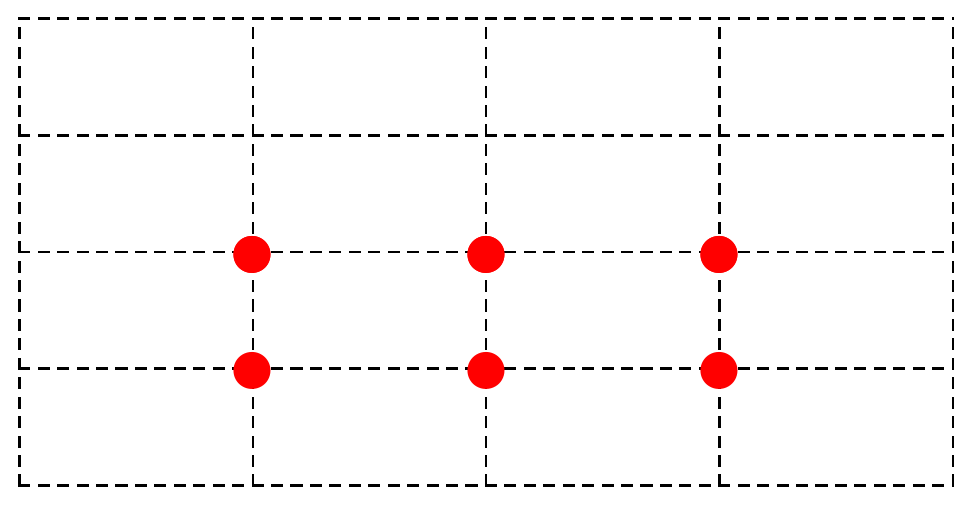} 
\end{minipage}
\quad
\begin{minipage}[b]{0.2\linewidth}
\includegraphics[width=\linewidth]{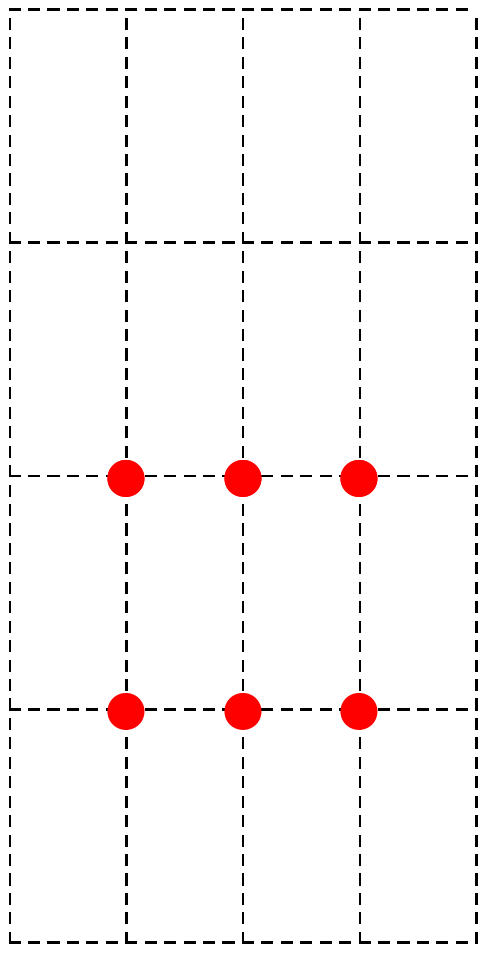} 
\end{minipage}
\quad
\begin{minipage}[b]{0.2\linewidth}
\includegraphics[width=\linewidth]{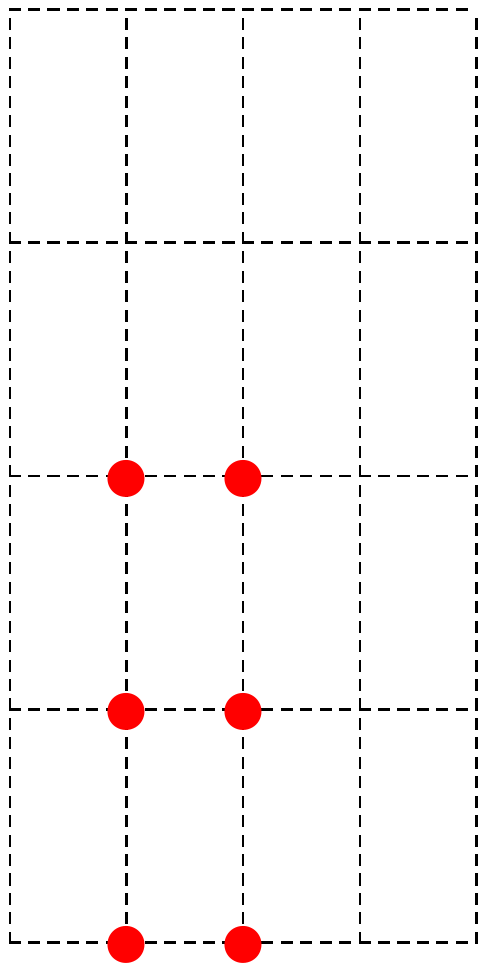} 
\end{minipage}
\caption{\label{fig-k_space_transform} The left panel shows the Fermi sea for 6 particles at $\tau=0.5i$. The new Fermi sea under the transformations prescribed by Eq.~\ref{k1} (Eq.~\ref{modtrans2}) is shown in the middle (right) panel. The Fermi seas  of the left panel and the right panel have the same shapes under a rotation by ninety degrees, while the Fermi sea of the middle panel has a different shape.  
}
\end{figure*}

As mentioned in the main text, the geometry of torus is parameterized by the modular parameter $\tau$. The correspondence between $\tau$ and the geometry is not one-to-one, and any physical observables should not depend on the parametrization. Under the modular transformation group formed by $\mathcal{T}:\tau\rightarrow \tau+1$, $\mathcal{S}:\tau\rightarrow -1/\tau$ and any combination of these two transformations \cite{Gunning62}, the mixing of a degenerate set of wave functions is expected to be closed.
The COM projected states transform under the modular transformation as follows: 
\be
\label{COM-transform}
P_{M} \Psi \equiv \Psi^{M} 
\rightarrow P'_{M}  \Psi \equiv \sum_{M'}K_{MM'} \Psi^{M'}
\ee
where $P_M$ is the projection operator into the COM momentum sector $M$. The matrix $K$ is a unitary matrix, which acts on a vector with the COM projected wave functions as its entries. The wave functions $\{\Psi^M\}$ are closed under modular transformation and the expectation value of any operator remains invariant. The wave functions that satisfy this property are said to be modular covariant \cite{Fremling14,Fremling19,Pu20}.

In this Appendix, we show the CF-BCS wave function Eq.~\ref{CF-BCS} is modular covariant both before LLL projection and after ``direct" LLL projection.

Let us begin by carefully defining the modular transformation for the CF Fermi sea and the CF-BCS wave functions that involve plane waves. For this purpose, we define the coordinates as $z=L(\tilde{x}+\tau \tilde{y})$, where $\tilde{x}$ and $\tilde{y}$ are the reduced coordinates along $L$ and $L\tau$ directions, respectively.  For $\phi_1=\phi_{\tau}=0$, $\vec{k}$ can be represented as
\begin{equation}
\label{k-vec}
k_{(m,n)}=m\left({2\pi\over L}\right)+ i{2\pi\over L}\left(\frac{n}{\tau_2}-\frac{m \tau_1}{\tau_2}\right)
\end{equation}
where $m,n$ are integers. For $\mathcal{T}$ transformation, both the physical coordinates $\vec{r}_i$ and wave vector $\vec{k}_j$ are invariant under modular transformation, and hence the Pfaffian part is also invariant. The $\mathcal{S}$ transformation requires some care. Under $\mathcal{S}$ transformation, even though the lattice remains invariant, the coordinates and other parameters transform as
\begin{equation}
z \rightarrow z'={|\tau| \over \tau}z; L \rightarrow L|\tau|; \phi_1 \rightarrow \phi_\tau; \phi_\tau \rightarrow -\phi_1
\label{modtrans1}
\end{equation}
 A direct evaluation of $\vec{k}_j$ using Eq.~\ref{k-vec} gives
\begin{equation}
\label{k1}
k_{(m,n)} \rightarrow k'_{(m,n)} = i\frac{|\tau|}{\tau}\Bigg(\frac{L}{N_{\phi}} \Bigg )(m + n\tau )
\end{equation}
However, this transformation produces an essentially different Fermi sea as shown in Fiq.~\ref{fig-k_space_transform}. The transformation that preserves the Fermi sea under the $\mathcal{S}$ transformation is
\begin{equation}
k \rightarrow k'={|\tau| \over \tau} k\;.
\label{modtrans2}
\end{equation}
Eqs. \ref{modtrans1} and \ref{modtrans2} define the $\mathcal{S}$ transformation.

We first demonstrate modular covariance for the ``unprojected" CF-BCS-paired wave function in Eq.~\ref{CF-BCS}. This wave function is made up of two parts, the Pfaffian of paired plane waves and the Laughlin wave function.  Simultaneous change of $z$ and $k$ ensures that  $e^{i\vec{k}\cdot (\vec{r}_i-\vec{r}_j)}$ remains invariant. On the other hand, $g_{\vec{k}}$ is just multiplied by an overall phase ${|\tau|\over \tau}$. Hence, the Pfaffian part remains invariant. It was shown in Reference ~\cite{Fremling19} that the Laughlin wave function is modular covariant. Therefore the unprojected CF-BCS-paired wave function is modular covariant. 

It is also straightforward to see the direct LLL projected wave function is modular covariant. The direct LLL projection operator can be written as $P_{\rm LLL}=\Pi_{n=1}^{\infty}\left(1-{a^\dagger a\over n}\right)$, in which $a^\dagger$ and $a$ are the ladder operators. Since $a^\dagger a$ is modular invariant, $P_{\rm LLL}$ commutes with the modular transformation (up to gauge transformation) \cite{Fremling19}. This proof does not extend to the JK projected CF-BCS wave function.

Numerical verification of modular covariance can be performed by testing Eq.~\ref{COM-transform}. 
Refs.~\cite{Fremling19,Pu20} have shown that, for a modular covariant wave function, Eq.~\ref{COM-transform} is satisfied with
\beq
K=\frac{1}{2}
\begin{bmatrix}
1 & 1\\
1 & -1
\end{bmatrix}
\eeq
This $K$ matrix can be obtained by noticing that the right-hand-side and left-hand-side of Eq.~\ref{COM-transform} are eigenstates of $t_{\rm CM}\left(L\tau/N_\phi\right)$ and $t_{\rm CM}\left(L/N_\phi\right)$ respectively. Thereby one can do a basis transformation from one to the other to derive the matrix elements.
We have confirmed numerically that the ratio of the wave functions on right-hand-side and the left-hand-side of Eq.~\ref{COM-transform} with the above $K$ remains a constant for different real space configurations,
for both the unprojected and the direct projected CF-BCS wave functions; this 
numerically confirms their equality modulo an overall normalization factor, and thus proves that these wave functions are modular covariant.  Alternatively, one can directly calculate the normalized overlap matrix $U_{MM'}=\bra{\mathcal{S}\Psi^{M}}\ket{\Psi^{M'}}$, where $\Psi^{M}$ refers to the wave function at COM momentum $M$ and $\mathcal{S}\Psi^{M}$ refers to the wave function after transformation. If the wave function is modular covariant, the overlap matrix should be unitary. We have found that the overlap matrix satisfies the relation $U^{\dagger}U=I$ (within statistical error) for both the unprojected and the direct projected CF-BCS wave functions (we have tested this for systems with up to 12 and 4 particles for the unprojected and direct projected wave functions, respectively).

\begin{figure}[t]
\includegraphics[width=0.9\columnwidth]{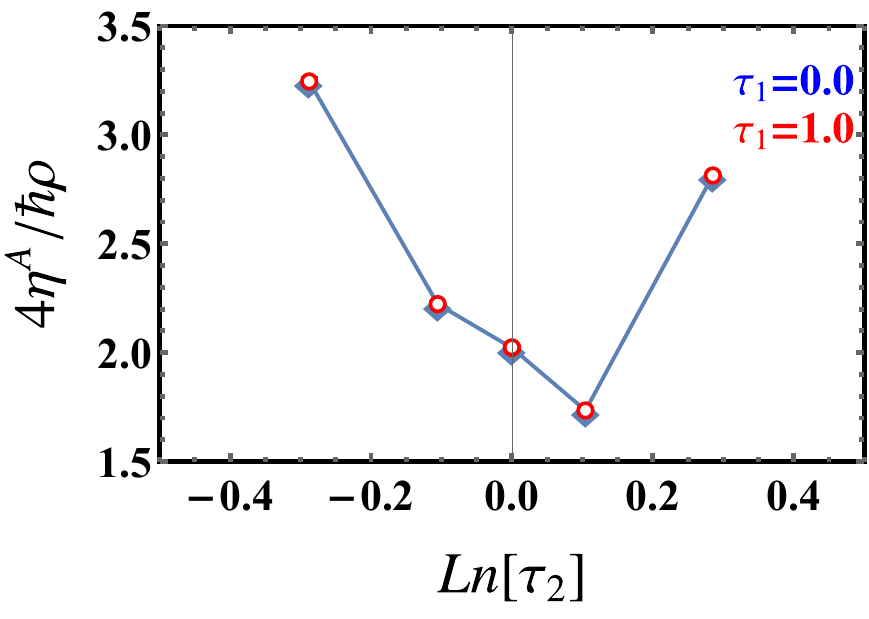} 
\caption{\label{S-trans}Hall viscosity $\eta^A$, quoted in units of $\hbar\rho/4$, for the JK projected CF-BCS wave function for different values of $\tau$. The ${\mathcal T}$ transformation $\tau\rightarrow \tau+1$ relates the red and the blue points at a given $\tau_2$, whereas the ${\cal S}$ transformation  $\tau\rightarrow -1/\tau$ relates $\ln \tau_2$ to $-\ln \tau_2$. Clearly, the JK wave function is covariant under the 
${\mathcal T}$ transformation, but not under the 
$\mathcal{S}$ transformation. The calculations are performed for a system with 12 particles with $\delta_{k_F}=0.5$.
}
\end{figure}

The modular covariance of the modified JK projection for Jain states and CFFS was proven in Ref.~\cite{Fremling19}. However, numerical tests show that the modified JK projected CF-BCS-paired wave function is {\it not} modular covariant: the JK projected wave function remains covariant under the $\mathcal{T}$ transformation but not for the $\mathcal{S}$ transformation. In Fig.~\ref{S-trans}, we show that the Hall viscosity calculated  for wave functions for values of $\tau$ related by $\mathcal{T}$ transformation are same. However, the Hall viscosity calculated for $\tau$ related by $\mathcal{S}$ transformation are not equal, which is evident from the asymmetry about $ln(\tau_2)=0$ line in the plot. 

\begin{figure}[t]

\includegraphics[width=0.9\columnwidth]{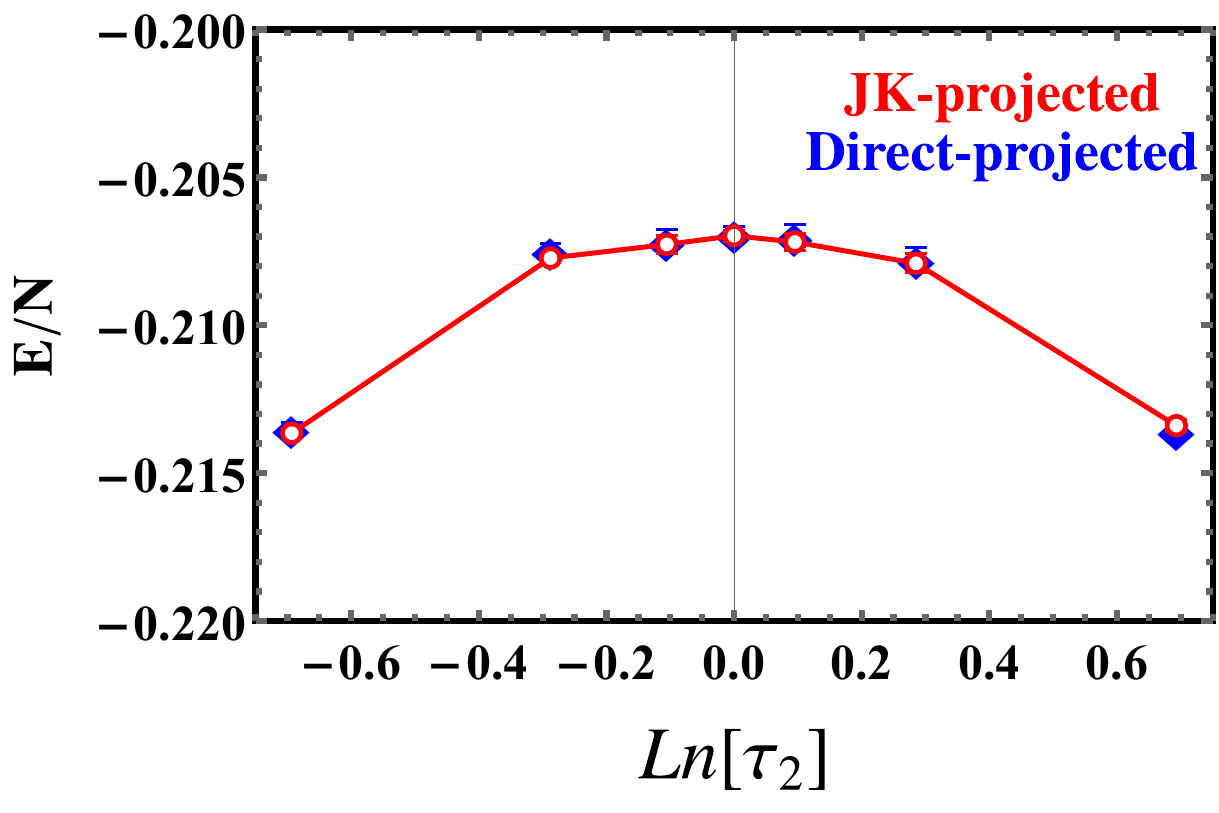} 
\caption{\label{Direct-results}Plot of the Coulomb  energy per particle in the LLL(in units of $e^2/\epsilon \ell$) for the direct projected and the JK projected CF-BCS wave functions for different values of $\tau$ related by $\mathcal{S}$ transformation. The blue and the red points represent the energies of the direct projected and JK projected CF-BCS wave functions, respectively. While the direct projected wave function is modular covariant, the JK projected wave function is not. Nonetheless, the energies of these wave functions are almost identical. 
Furthermore, the energies are also symmetric, within numerical error, with respect to the line $ln(\tau_2)=0$, which is consistent with covariance under the $\mathcal{S}$ transformation. The energies do not include the self interaction energy. 
The calculations are performed for a system with 4 particles for $\delta_{k_F}=0.5$ in the full Brillouin zone.   
}
\end{figure}

{Fortunately, even though the JK projected wave function is, strictly speaking, not modular covariant, it produces energies very close to that of the direct projected wave function, which is modular covariant. This is demonstrated in Fig.~\ref{Direct-results}, which shows the LLL energy for the JK-projected wave functions related by  the ${\cal S}$ transformation. 
This suggests that modular transformation preserves, to a good approximation, for the absolute value of the JK-projected wave function, and justifies the use of the JK-projected CF-BCS wave function in our variational study. We note here that the difference between the energies of the JK-projected and the direct projected BCS wave functions, which is less than $0.001$ per particle, is much smaller than the energy gain as a function of $\delta_{k_F}$, which is on the order of 0.008 per particle for $\delta_{k_F}\sim 1.0$.}

\section{A brief review of lattice Monte Carlo}

The lattice Monte Carlo approach used in this paper was originally proposed by Wang {\it et al.} ~\cite{Wang19}. Here we provide a review of this approach giving further details with a slightly different logical organization. For notational facility, we only derive the result for single particle operator $\sum_i O(\vec{r}_i)$. The generalization to a two-body operator $\sum_{i<j} O(\vec{r}_i-\vec{r}_j)$ is straightforward. We use $\vec{x}$ to represent $(\vec{r}_1,\vec{r}_2,\dots \vec{r}_N)$ and omit the subscript $i$ in $O(\vec{r}_i)$ for simplicity.

The object to calculate is:
\be
{\langle \Psi_1|O(\vec{r})|\Psi_2\rangle\over \sqrt{\langle \Psi_1|\Psi_1\rangle\langle \Psi_2|\Psi_2\rangle}}={\int d^2x\Psi_1^*(\vec{x})O(\vec{r})\Psi_2(\vec{x})\over \sqrt{\langle \Psi_1|\Psi_1\rangle\langle \Psi_2|\Psi_2\rangle}}
\ee
in a periodic geometry. The aim is to replace the continuous integral by a summation on discrete lattice points $\vec{x}_i=(m_i\vec{L}_1+n_i\vec{L}_2)/N_\phi$: 
\be
\label{Lat}
{\langle \Psi_1|O(\vec{r})|\Psi_2\rangle\over \sqrt{\langle \Psi_1|\Psi_1\rangle\langle \Psi_2|\Psi_2\rangle}}={\sum'_{\vec{x}}\Psi_1^*(\vec{x})O^{\rm Lat}(\vec{r})\Psi_2(\vec{x})\over \sqrt{\sum'_{\vec{x}} |\Psi_1|^2\sum'_{\vec{x}} |\Psi_1|^2}}
\ee
The central result is to derive $O^{\rm Lat}(\vec{r})$ given any $O(\vec{r})$ when $\Psi_1$ and $\Psi_2$ are confined in the $n$th LL. Note that we have written the normalization factor explicitly, because Monte Carlo automatically includes the normalization factors, which are different for continuous space and discrete space.

For later reference, we list the Fourier transformations for both continuous and discrete spaces:
\be
O(\vec{r})={1\over 2\pi N_\phi}\sum_{\vec{q}}O(\vec{q})e^{i\vec{q}\cdot\vec{r}}
\ee
\be
O(\vec{q})=\int d^2r O(\vec{r})e^{-i\vec{q}\cdot\vec{r}}
\ee
\be
O^{\rm Lat}(\vec{r})={1\over 2\pi N_\phi}\sum'_{\vec{q}}O^{\rm Lat}(\vec{q})e^{i\vec{q}\cdot\vec{r}}
\ee
\be
O^{\rm Lat}(\vec{q})={2\pi\over N_\phi}\sum'_{\vec{x}} O^{\rm Lat}(\vec{r})e^{-i\vec{q}\cdot\vec{r}}
\ee
Here $\sum'_{\vec{q}}$ refers to summation within the first BZ, $\sum_{\vec{q}}$ to summation over the whole $\vec{q}$ space, and $\sum'_{\vec{x}}$ to summation over lattice points in the principal region of torus.

Let us first derive a useful relation (Eq.~\ref{qR}) based on the above equations: 
\ba
\label{qR}
\int d^2x\Psi_1^*(\vec{x})e^{i\vec{q}\cdot\vec{r}}\Psi_2(\vec{x})&=&f_n(q)\langle \Psi_1|e^{i\vec{q}\cdot\vec{R}}|\Psi_2\rangle \nonumber\\ 
2\pi N_\phi\Psi_1^*(\vec{x})\Psi_2(\vec{x})&=&\sum_{\vec{q}}e^{-i\vec{q}\cdot\vec{r}}f_n(q)\langle \Psi_1|e^{i\vec{q}\cdot\vec{R}}|\Psi_2\rangle \nonumber\\
2\pi N_\phi\Psi_1^*(\vec{x})\Psi_2(\vec{x})&=&\sum'_{\vec{q}}\left[f_n(q)\right]_{N_\phi}e^{-i\vec{q}\cdot\vec{r}}\langle \Psi_1|e^{i\vec{q}\cdot\vec{R}}|\Psi_2\rangle\nonumber\\
{2\pi\over N_\phi}\sum'_{\vec{r}}\Psi_1^*(\vec{x})\Psi_2(\vec{x})e^{i\vec{q}\cdot\vec{r}}&=&\left[f_n(q)\right]_{N_\phi}\langle \Psi_1|e^{i\vec{q}\cdot\vec{R}}|\Psi_2\rangle
\ea
In the first line we used $e^{i\vec{q}\cdot\vec{r}}=f_n(q)e^{i\vec{q}\cdot\vec{R}}$ for the $n$th LL, in which $f_n(q)=e^{-{|q|^2\over 4}}L_n\left({|q|^2\over 2}\right)$ is the form factor ($L_n(x)$ is the Laguerre polynomial). In the third line, we used $e^{i\vec{q}\cdot\vec{R_j}}=t_j(iq)$ and the periodicity of $|\Psi_2\rangle$ and $e^{-i\vec{q}\cdot\vec{r}}$ when $\vec{q}=N_\phi\vec{q}'$ and $\vec{r}$ is on a lattice point. Here, $\left[f_n(q_{l,m})\right]_{N_\phi}=\sum_{\vec{q}'_{j,k}}f_n(q_{l,m}+N_\phi q'_{j,k})e^{i(-k\phi_1+j\phi_2)}\times (-1)^{lk-mj+N_\phi jk}$ is called the compacitified form factor.  Line four comes from the Fourier transform on lattice points. Eq.~\ref{qR} enables us to evaluate $e^{i\vec{q}\cdot\vec{R}}$ on discrete lattice points. 

Note that in the above derivation we actually consider only one space coordinate, which is $\vec{r}$, while treating those $\vec{r}_i\neq \vec{r}$ as parameters.. To get the full inner product on the right hand side, we need to do integration over all space coordinates. Let us consider $\Psi(r_1,r_2\dots r_N)=\psi_1(r_1)\psi_2(r_2)\dots \psi_N(r_N)$, i.e. a product state. Taking $\vec{q}=0$ in Eq.~\ref{qR}, we find:
\ba
{2\pi\over N_\phi}\sum'_{\vec{r_1}}\psi_1^*(\vec{r_1})\psi_2(\vec{r_1})
=\left[f_n(0)\right]_{N_\phi}\langle \psi_1|\psi_2\rangle
\ea
For the full inner product of two N-particle states, the final form should be (note that on the left hand side it is now $\sum'_{\vec{x}}$):
\ba
\label{qR2}
&&\left({2\pi\over N_\phi}\right)^N\sum'_{\vec{x}}\Psi_1^*(\vec{x})\Psi_2(\vec{x})e^{i\vec{q}\cdot\vec{r}}\\ \nonumber
&=&\left[f_n(q)\right]_{N_\phi}\left[f_n(0)\right]_{N_\phi}^{N-1}\langle \Psi_1|e^{i\vec{q}\cdot\vec{R}}|\Psi_2\rangle
\ea
A general many-particle wave function can be expanded as a summation over product states, so the above derivation still holds.

For a general one-body operator $O(\vec{r})$:
\begin{widetext}
\ba
\label{Lator}
\langle \Psi_1|O(\vec{r})|\Psi_2\rangle&=&{1\over 2\pi N_\phi}\sum_{\vec{q}}O(\vec{q})\langle \Psi_1|e^{i\vec{q}\cdot\vec{r}}|\Psi_2\rangle \nonumber\\
&=&{1\over 2\pi N_\phi}\sum_{\vec{q}}O(\vec{q})f_n(q)\langle \Psi_1|e^{i\vec{q}\cdot\vec{R}}|\Psi_2\rangle \nonumber\\
&=&{1\over 2\pi N_\phi}\sum'_{\vec{q}}O^{GC}(\vec{q})\langle \Psi_1|e^{i\vec{q}\cdot\vec{R}}|\Psi_2\rangle \nonumber\\
&=&\sum'_{\vec{x}}\Psi_1^*(\vec{x})\Psi_2(\vec{x})\left({2\pi\over N_\phi}\right)^N\left[f_n(q)\right]^{-1}_{N_\phi}\left[f_n(0)\right]_{N_\phi}^{-(N-1)}{1\over 2\pi N_\phi}\sum'_{\vec{q}}O^{GC}(\vec{q})e^{i\vec{q}\cdot\vec{r}}
\ea
\end{widetext}
In the 3rd line we used the periodicity and $O^{GC}(\vec{q})=\sum_{\vec{q}'}O(\vec{q}+N_\phi\vec{q}')f_n(q+N_\phi q')$. In the 4th line we used Eq.~\ref{qR2}. For the special case $O(\vec{r})=1$, Eq.~\ref{Lator} becomes:
\be
\label{Lat1}
\langle \Psi_1|\Psi_2\rangle=\sum'_{\vec{x}}\Psi_1^*(\vec{x})\Psi_2(\vec{x})\left({2\pi\over N_\phi}\right)^N\left[f_n(0)\right]_{N_\phi}^{-N}
\ee
Finally, plugging Eq.~\ref{Lator} and Eq.~\ref{Lat1} into Eq.~\ref{Lat}, we get:
\be
O^{Lat}(\vec{r})={1\over 2\pi N_\phi}\sum'_{\vec{q}}{\left[f_n(0)\right]_{N_\phi}\over \left[f_n(q)\right]_{N_\phi}}O^{GC}(\vec{q})e^{i\vec{q}\cdot\vec{r}}
\ee
An analogous treatment for a two-body operator $O^{Lat}(\vec{r}_i-\vec{r}_j)$ produces:
\be
\label{Lat2b}
O^{Lat}(\vec{r}_i-\vec{r}_j)={1\over 2\pi N_\phi}\sum'_{\vec{q}}\left({\left[f_n(0)\right]_{N_\phi}\over \left[f_n(q)\right]_{N_\phi}}\right)^2O^{GC}(\vec{q})e^{i\vec{q}\cdot(\vec{r}_i-\vec{r}_j)}
\ee
and $O^{GC}(\vec{q})=\sum_{\vec{q}'}O(\vec{q}+N_\phi\vec{q}')f^2_n(q+N_\phi q')$ for a two-body operator.

Eq.~\ref{Lat2b} corresponds to Eq.(17) in Ref.~\cite{Wang19}. 

\section{Overlaps with the exact ground state}
\label{ex-overlap}
The exact ground states are calculated using the DiagHam package. They are written in the Fock space with basis $\ket{k_1,k_2,..,k_N}$ where $k _i=0,...,N_{\phi}-1$ represents the occupied orbital number. For fermions, the $k _i$s are arranged in ascending order i.e. $k_i< k_j$ for $i<j$. The single particle orbitals in the LLL can be written as \cite{Pu20}
\beq
\psi(z,\bar{z}) = \mathcal{N}e^{\frac{z^2-|z|^2}{4\ell^2}}f^{(k)}_0(z,\bar{z})
\eeq
where
\beq
f^{(k)}_0(z,\bar{z}) =  \vartheta \begin{bmatrix} {\frac{k} {N_{\phi}}+\frac{\phi_{1}}{2 \pi N_{\phi}}} \\ -\frac{\phi_2}{2 \pi} \end{bmatrix} \Bigg( \frac{N_{\phi}z}{L} \Bigg | N_{\phi}z \Bigg ).
\eeq
$k=0,...,N_{\phi}-1$ represents the momentum of the state under the translation by $t(L/N_{\phi})$. The normalization with respect to the physical coordinates can be written as $\mathcal{N}=1/\sqrt{\ell L \sqrt{\pi}}$. Obviously, the basis $\ket{k_1,k_2,..,k_N}$ are eigenstates of the center-of-mass magnetic translation $t_{\rm CM}\left(L/N_\phi\right)$.

The relative magnetic translation operator is defined as 
\beq
\tilde{t}_i(a) = t_i(a)\prod _{j=1}^N t_j(-a/N) 
\eeq
The basis $\ket{k_1,k_2,..,k_N}$ are automatically eigenstates of $\tilde{t}_i(L)$ [here we only consider $\nu=1/2$, for which GCD($N, N_\phi$)$=N$] as we choose the primary axis along the $x$ direction:
\beq
\tilde{t}_i(L ) \ket{k_1,k_2,..,k_N} = e^{i\frac{2 \pi }{N}\sum_{i=1} ^N k_i}  \ket{k_1,k_2,..,k_N} \nonumber \\
\eeq
where the total momentum is defined as $\sum_{i=1} ^N k_i$(mod $	N$). The momentum sector $k_x$ for the state can be obtained from the relation:
\beq
e^{-i\frac{k_xL_x}{N}}=e^{i2\pi \sum_{i=1}^N k _i /N} 
\eeq
However, the basis $\ket{k_1,k_2,..,k_N}$ are not eigenstates of the magnetic translation in the other direction $\tilde{t}_i(L \tau)$
\beq
\label{rel-tran}
\tilde{t}_i(L \tau) \ket{k_1,k_2,..,k_N} = \ket{k_1+2,..,k_N+2}. \nonumber \\
\eeq
After applying $\tilde{t}_i(L \tau)$ operator $Z$ times, we go over $Z$ different basis states $\ket{k_1+2s,k_2+2s,..,k_N+2s}$ with $s=0,1,\cdots Z-1$ and finally get back to the original state, because the momentum is defined mod $N_\phi$. The eigenstates of $\tilde{t}_i(L \tau)$ are obtained by taking superposition of these states \cite{Bernevig12}.

\beq
\label{orbit}
\ket{k_x,k_y(n)}= \sum_{s=0}^{Z-1} e^{i\frac{2\pi n}{N}s} \ket{k_1+2s,k_2+2s,..,k_N+2s} \nonumber \\
\eeq
where $n \in [0,...,N-1]$ is an integer which determines $k_y$: 
\beq
\tilde{t}_i(L \tau)\ket{k_x,k_y}=e^{i2\pi n /N}\ket{k_x,k_y}=e^{-i\frac{k_yL_y}{N}}\ket{k_x,k_y}.\nonumber\\
\eeq
The momentum sectors are specified by the eigenvalues of $t_{\rm CM}\left(L/N_\phi\right)$, $\tilde{t}_i(L)$, and $\tilde{t}_i(L \tau)$.

Suppose the dimension of the momentum sector that our trial state belongs to is $D$. We name the basis states as $\ket{\phi_m}$, which are eigenstates of $t_{\rm CM}\left(L/N_\phi\right)$, $\tilde{t}_i(L)$, and $\tilde{t}_i(L \tau)$. To calculate the overlap of the trial states with the exact states, we need to decompose the trial wave function\cite{Sreejith18}
\beq
\ket{\Psi_{\rm trial}}=\sum_{m=1}^D c_m \ket{\phi _m}
\eeq
This is done by choosing $D'$ sets of real space configurations $r_1^{(\alpha)}, r_2^{(\alpha)},\cdots r_N^{(\alpha)}$ where $\alpha=1,2,\cdots D'$ and solve the linear equations:
\be
\label{inverse}
\Psi_{\rm trial}\left[r_i^{(\alpha)}\right]=\sum_{m=1}^D c_m \phi _m \left[r_i^{(\alpha)}\right] \quad \alpha=1,2,\cdots D'
\ee
The system of linear equations are solved using least square method, in which one minimizes $\sum_{\alpha=1}^{D'} |\Psi_{\rm trial}\left[r_i^{(\alpha)}\right]-\sum_{m=1}^D c_m \phi _m \left[r_i^{(\alpha)}\right]|^2$ to find $c_m$. We used the linear algebra package in Scipy to solve the system of equations. The system of equations in Eq.~\ref{inverse} can be ill conditioned for certain sets of configurations and the solution can be unstable for such system; i.e. , a small variation in the values of $\Psi_{\rm trial}[r_i]$ can lead to huge differences in the values of the solution $c_m$'s. For our calculation, we use an over-determined  system of equations i.e. we consider more equations than the number of coefficients (i.e. $D'>D$). This is done to ensure the numerical stability of the solution\cite{Balram21}.
 Further, in order to improve the conditioning of the matrix $\phi _m \left[r_i^{(\alpha)}\right]$, we used particle configurations obtained after Monte Carlo thermalization. 
\end{appendix}


\end{document}